\documentclass{aa}  

\usepackage{graphicx}
\usepackage[varg]{txfonts}
\usepackage{natbib}
\usepackage{amsmath}
\usepackage{xcolor}
\usepackage[normalem]{ulem}
\usepackage{color}
\usepackage{xcolor}
\definecolor{xlinkcolor}{cmyk}{1,1,0,0}
\usepackage[bookmarks=true,         
bookmarksopen=true,
pdfnewwindow=true,      
colorlinks=true,    
linkcolor=xlinkcolor,     
citecolor=xlinkcolor,     
filecolor=xlinkcolor,  
urlcolor=xlinkcolor,      
final=true]{hyperref} 
\hypersetup{pdfstartview=Fit}
\usepackage{bookmark}

\begin{document}

	\title{Improved asteroseismic inversions for red-giant surface rotation rates}
	
	\author{F.\ Ahlborn\inst{1}
		\and E.\ P.\ Bellinger\inst{1,2}
		\and S.\ Hekker\inst{3,4,2}
		\and S.\ Basu\inst{5}
		\and D.\ Mokrytska\inst{3,4,2}}
	\institute{Max-Planck-Institut für Astrophysik, Karl-Schwarzschild-Stra\ss e 1, 85748 Garching, Germany\\
		email: \url{fahlborn@mpa-garching.mpg.de}
		\and Stellar Astrophysics Centre, Department of Physics and Astronomy, Aarhus University, Ny Munkegade 120, DK-8000 Aarhus C, Denmark
		\and Center for Astronomy (ZAH/LSW), Heidelberg University, Königstuhl 12, 69117 Heidelberg, Germany
		\and Heidelberger Institut für Theoretische Studien, Schloss-Wolfsbrunnenweg 35, 69118 Heidelberg, Germany
		\and Department of Astronomy, Yale University, New Haven, CT 06520, USA
	}
	
	\date{Received ; accepted }
	
	\abstract
	{Asteroseismic observations of internal stellar rotation have indicated a substantial lack of angular momentum transport in theoretical models of subgiant and red-giant stars. Accurate core and surface rotation rate measurements are therefore needed to constrain internal transport processes included in the models.}
	{We eliminate substantial systematic errors of asteroseismic surface rotation rates found in previous studies.}
	{We propose a new objective function for the Optimally Localized Averages method of rotational inversions for red-giant stars, which results in more accurate envelope rotation rate estimates obtained from the same data. We use synthetic observations from stellar models across a range of evolutionary stages and masses to demonstrate the improvement.}
	{We find that our new inversion technique allows us to obtain estimates of the surface rotation rate that are independent of the core rotation. For a star at the base of the red-giant branch, we reduce the systematic error from about 20\% to a value close to 0, assuming constant envelope rotation. We also show the equivalence between this method and the method of linearised rotational splittings.}
	{Our new rotational inversion method substantially reduces the systematic errors of red-giant surface rotation rates. In combination with independent measures of the surface rotation rate, this will allow better constraints to be set on the internal rotation profile. This will be a very important probe to further constrain the internal angular momentum transport along the lower part of the red-giant branch.}
	
	
	\maketitle
	%
	
	\section{Introduction} 
	The observations of internal rotation rates in evolved stars have revealed a substantial mismatch with respect to theoretical models. These models predict the core to rotate orders of magnitude faster than is observed. The mismatch between theory and observations is indicative of a lack of angular momentum transport from the core to the envelope in the models \citep{eggenberger2012,ceillier2013,marques2013,spada2016,eggenberger2017,ouazzani2019,fuller2019}. To remedy the lack of angular momentum transport in later evolutionary stages, different effects have been considered so far, e.g., the effects of magnetic fields \citep{cantiello2014,fuller2019}, internal gravity waves \citep{fuller2014} and mixed modes \citep{belkacem2015a,belkacem2015b}. To set more constraints on the internal mixing processes and further understand the angular momentum transport mechanisms active in stars, it is crucial to accurately measure core and envelope rotation rates in red giants.
	
	In recent years, red giants have been shown to be an excellent testbed of stellar astrophysics owing to the detection of solar-like oscillations in time series of their surface brightness variations. In these stars, oscillations are stochastically excited by turbulent convection in the outer layers which propagate through the stellar interior. In red giants, so-called mixed modes are observed which have a gravity (g) component in the core and a pressure (p) component in the envelope \citep[e.g.,][]{beck2011,bedding2011,mosser2011,mosser2014}. Rotation splits oscillation modes with a spherical degree larger than zero into prograde, retrograde and standing components. These components oscillate at different frequencies where the difference in frequency, so-called rotational splitting, depends on the rotation of the part of the star to which the oscillation mode is sensitive. Measuring these rotational splittings as a function of frequency allows us to draw conclusions about the internal radial rotation profile. \cite{beck2012} have shown that in order to explain the observed rotational splittings of a red giant, the core of the star needs to rotate at least 10 times faster than the envelope. To date, a number of methods have been suggested to measure the internal rotation in stars. \cite{mosser2012} have shown that the rotational splittings of g-dominated modes constrain the core rotation rate of red giants. This approach has been further developed by \cite{goupil2013} who used the so-called mode trapping---the ratio between core and total mode inertia---to estimate the sensitivity of the rotational splittings to different regions in the stellar interior. They show that the splittings can be approximated as a linear function of the mode trapping. The slope and the intercept of this relation allow us to estimate the average core and envelope rotation rates. \cite{deheuvels2015} applied this method to measure the internal rotation in a set of core helium-burning stars. To probe the internal rotation as a function of depth, \cite{beck2014} proposed a forward modelling approach \citep[see also][]{beck2018}.
	
	Rotational splittings can further be analysed using a linear perturbative approach, allowing localised estimation of the internal rotation rates. Such an approach is known as a rotational inversion and has been applied very successfully to measure the internal rotation profile of the Sun \citep[e.g., ][]{schou1998,christensen1990,howe2009}. Using the high precision space photometry obtained with the \textit{Kepler} telescope \citep{borucki2010}, it became possible to compute rotational inversions for red-giant stars \citep{deheuvels2012,deheuvels2014,dimauro2016,dimauro2018,triana2017}. Using rotational inversions, the core rotation rates of red giants can be computed with relatively high accuracy. The determination of rotation at intermediate radii remains impossible with low-degree modes \citep{ahlborn2020}. 
	
	Recently, \cite{ahlborn2020} have shown that estimated surface rotation rates suffer from substantial systematic errors. This effect becomes especially pronounced for more evolved stars along the red-giant branch (RGB). The systematic errors arise because the estimated surface rotation rates show a substantial amount of sensitivity to the core region. As the core of red-giant stars is generally rotating much faster than the envelope, this leads to an overestimation of the surface rotation rate. 
	
	In this paper, we show that the occurrence of the core sensitivity of the estimated surface rotation rate can be attributed to the specific choice of the inversion method. We propose an extension to the objective function in the commonly-used Multiplicative Optimally Localised Average (MOLA) inversions method \citep{backus1968} that we call eMOLA. To demonstrate and assess our method, we apply it to stellar models and synthetic data along the lower part of the RGB up to the bump in luminosity. We find that the new objective function allows us to completely suppress any sensitivity to the core region in the estimated surface rotation rates. Finally, we compare the rotational inversion results using this new objective function to results from the linear splittings approximation (LSA) by \cite{goupil2013}. Using linear regression, we estimate the  core and surface rotation rates for the LSA, and construct sensitivity kernels for both estimates. We demonstrate that they are very similar to the averaging kernels obtained from eMOLA rotational inversions. This explains why the LSA recovers input rotation rates with similar accuracy as the eMOLA rotational inversions. Therefore, both methods allow us to probe both the core and envelope rotation rates along the lower part of the RGB with high accuracy.

	\section{Rotational inversions}
	\label{secMOLA}
	Oscillation modes with spherical degree $l>0$ are degenerate in their azimuthal order. Stellar rotation lifts this degeneracy and splits oscillation frequencies into multiplets of $2l+1$ modes. The frequency difference between subsequent peaks of a multiplet is called rotational splitting, and is denoted by $\delta\omega$. For slowly rotating stars, it can be described as a linear perturbation to the mode frequency:
	\begin{align}
	\omega_{n,l,m}=\omega_{n,l}+m\,\delta\omega_{n,l},
	\end{align}
	where $n$ and $m$ denote the radial and azimuthal order of the mode, respectively. The rotational splitting can be computed as the integral over the internal rotation profile $\Omega(r)$ weighted with a so-called rotational kernel $\mathcal{K}_i(r)$:
	\begin{align}
	\delta\omega_{nl}=\int_0^R\mathcal{K}_{nl}(r)\,\Omega(r)\,\text{d}r,
	\label{eqsplittings}
	\end{align}
	where $r$ refers to the radial coordinate \citep[e.g., ][]{gough1981}. To determine the internal stellar rotation rates at selected target radii $r_0$ we use a modified version of the MOLA inversion technique \citep{backus1968}. In MOLA inversions, so-called averaging kernels (AKs) are constructed by linearly combining the rotational kernels with inversion coefficients $c_i(r_0)$:
	\begin{align}
	K(r,r_0)=\sum_{i\in\mathcal{M}}c_i(r_0)\,\mathcal{K}_i(r),\label{eqavgkernel}
	\end{align}
	where $\mathcal{M}$ refers to the mode set of interest. The inversion coefficients are computed such that the AKs are as localised as possible at the target radius. Given a localised AK, the rotation rate at the target radius can be estimated as:
	\begin{align}
	\overline{\Omega}(r_0)=\int_{0}^{R}K(r,r_0)\,\Omega(r)\,\mathrm{d}r=\sum_{i\in\mathcal{M}}c_i(r_0)\,\delta\omega_i.
	\label{eqomegaavg}
	\end{align}
	As mentioned, \cite{ahlborn2020} have shown that surface AKs for red-giant stars are not well localised and show considerable sensitivity to the core rotation. This is a major difference to solar rotational inversions, for which MOLA inversions have been shown to work accurately \citep{christensen1990,schou1998}. For the Sun the rotational splittings of hundreds of pure p-modes are available. This allows for the construction of localised averaging kernels throughout large parts of the Sun. In contrast, the mode sets for red giants are limited to a small number of modes (on the order of ten), with limited spherical degree (so far only $l=1$) and most importantly all oscillation modes are mixed modes. This means that all modes used are sensitive to both the core and the envelope. Note, however, that the ratio of core to envelope sensitivity may vary. As a consequence, it is not possible to select a subset of modes which is sensitive to either the core or the envelope. Instead, the rotational inversion method has to be constructed such that it appropriately localises the averaging kernel. In the following, we will discuss the original formulation of MOLA inversions and subsequently a suitable modification of the MOLA inversion to mitigate the aforementioned problems.
	\subsection{Objective function}
	The inversion coefficients $c_i(r_0)$ are obtained by minimising an objective function. The MOLA objective function is given by 
	\begin{align}
	Z_\text{MOLA}=\int_0^RK(r,r_0)^2J(r,r_0)\,\text{d}r+\frac{\mu}{\mu_0}\sum_{i,j\in\mathcal{M}}c_i(r_0)c_j(r_0)E_{ij},
	\label{eqobjMOLA}
	\end{align}
	where  the function $J(r,r_0)$ is a function weighting the amplitude of the AK away from the target radius. A common choice is \citep[][]{backus1968,gough1985}
	\begin{align*}
	J(r,r_0)&=12(r-r_0)^2/R,
	\end{align*}
	where $R$ denotes the stellar radius. The parameter $\mu$ denotes the so-called trade-off parameter, which balances the uncertainty of the solution from propagated errors with the resolution of the inversions as indicated by the width of the AKs. The trade-off parameter is scaled with $\mu_0$  
	\begin{align}
	\mu_0&=\frac{1}{M}\sum_{i\in\mathcal{M}}E_{ii},
	\end{align}
	where $M$ denotes the number of modes and $\mathbf{E}$ is the error variance-covariance matrix from the observations. Therefore, the second term is independent of the absolute values of the uncertainties.
	\subsection{Extended objective function}
	\begin{figure}
		\centering
		\includegraphics{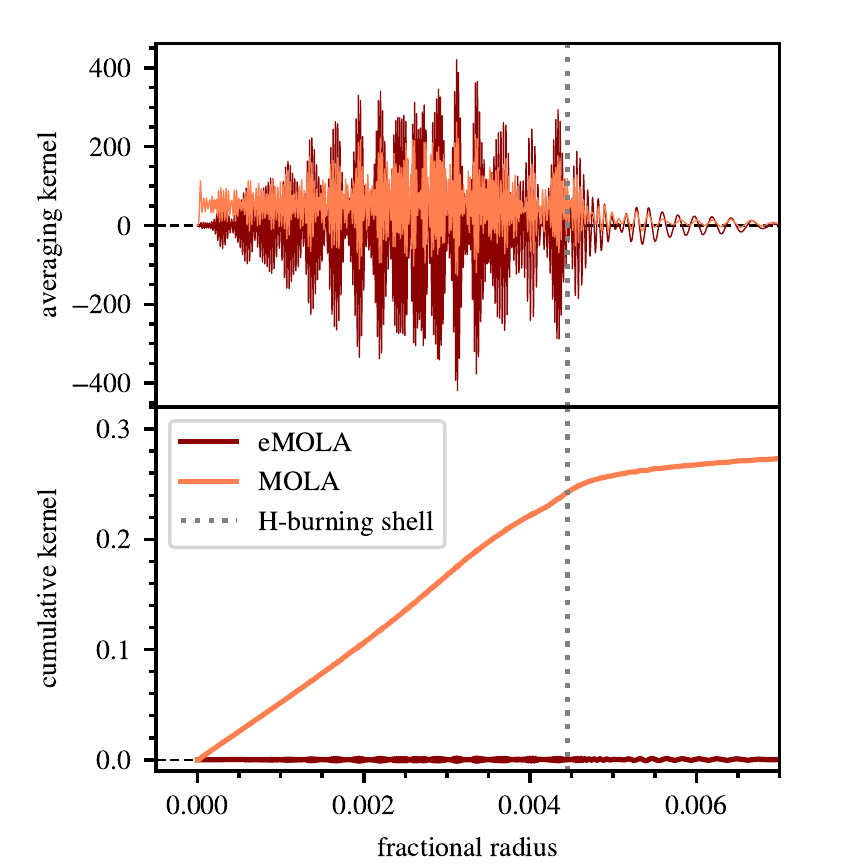}
		\caption{Comparison of surface averaging kernels obtained with the original MOLA and the extended MOLA objective functions. The AKs and the corresponding cumulative kernels are shown in the upper and lower panels, respectively. The grey dotted line indicates the position of the hydrogen burning shell. While the surface averaging kernel obtained with MOLA shows substantial sensitivity to the core, eMOLA is able to eliminate this sensitivity.}
		\label{figeMOLAcomparison}
	\end{figure}
	The original choice of the objective function (Eq.~\ref{eqobjMOLA}) ensures that the absolute value of the AK is as small as possible when moving away from the target radius. However, it does not penalise the accumulation of sensitivity in regions away from the target radius. This can be illustrated as follows: a surface averaging kernel (SAK) that accumulates sensitivity in the core with small amplitude would be favoured over a large-amplitude oscillatory solution without cumulative sensitivity to the core. In red giants, the oscillations in the core have very small spatial scales that are much smaller than those on which the internal rotation profile is expected to change. Therefore, when integrating Eq.~(\ref{eqomegaavg}) an oscillatory solution does not alter the estimated rotation rate as long as it is not accumulating sensitivity in the core. Summarising, an AK that suppresses cumulative sensitivity away from the target radius should be favoured over solutions with small amplitude but noticeable cumulative sensitivity. In Fig.~\ref{figeMOLAcomparison} two SAKs and the related cumulative kernels are shown which exemplify the two cases discussed above. The kernel labelled with MOLA has a comparatively small amplitude in the core, but it accumulates about 30\% of its sensitivity in these very deep layers. Due to the small-scale spatial oscillations in the g-mode component it is not possible to find linear combinations of rotational kernels, that completely remove the sensitivity to the core in the MOLA SAK. On the other hand, the kernel labelled with eMOLA has a comparatively high amplitude in the core, and does not accumulate any sensitivity in this region. In this SAK, subsequent small-scale spatial oscillations of the g-mode component cancel each other when integrating. A surface rotation rate estimated with the eMOLA AK is insensitive to the core region, while a surface rotation rate estimated with the MOLA kernel contains substantial sensitivity in the core region.
	
	To calculate the inversion coefficients for an AK with suppressed sensitivity (eMOLA), we therefore propose to modify the original MOLA objective function (Eq.~\ref{eqobjMOLA}). We add a term which penalises cumulative sensitivity in regions away from the target radius. We propose to include the penalised sensitivity in a quadratic way:
	\begin{align}
	Z=Z_\text{MOLA}+\theta\left[\int_0^RK(r,r_0)J(r,r_0)\,\text{d}r\right]^2
	\label{eqobj2}
	\end{align}
	where $\theta$ is a second trade-off parameter.
	
	The second term of Eq.~(\ref{eqobj2}) now accounts for any cumulative sensitivity away from the target radius. The sensitivity away from the target radius is weighted with the function $J(r,r_0)$ as done for the squared amplitude in the original term. The parameter $\theta$ controls the balance between the requirements of a small amplitude and no cumulative sensitivity away from the target radius. For $\theta=0$ the method is identical to the original MOLA description. For $\theta\to\infty$ the new term will dominate. A suitable value for the parameter $\theta$ has to be found by numerical experiments, as done for the trade-off parameter $\mu$. While the objective function of the extended MOLA inversion is different, the general principle of deriving the inversion coefficients is the same as for the original MOLA. This is shown in Appendix~\ref{secderiv}.
	
	\section{Extended MOLA rotational inversion results}
	\label{seceMOLAsingle}
	In this section, we apply the eMOLA inversion method to a set of synthetic data. We have constructed three different stellar evolution tracks with masses of 1.0, 1.5 and 2.0 $\text{M}_\odot$ with solar metallicity using Modules for Experiments in Stellar Astrophysics \cite[MESA version 8845,][and references therein]{paxton2019}. Using the stellar oscillation code GYRE \citep{townsend2013,townsend2018}, we compute the oscillation frequencies and associated rotational kernels. For the details of the stellar models we refer to \citealt[][see their Sec.~2 and Appendix~A.1]{ahlborn2020}. To compute the synthetic rotational splittings we use three different rotation profiles. We use two step-rotation profiles, one with a step at 1.5 times the radius of the hydrogen burning shell $r_\text{H}$, and one with a step at the base of the convective envelope $r_{\rm rcb}$. In the following, we will refer to these profiles as ``core step'' and ``envelope step'' profiles, respectively. In addition, we compute a profile that has a constant rotation below the base of the convective envelope and a power law decrease in the convective envelope. We will refer to this profile as the "convective power law" profile. For the details of the rotation profiles we refer to Appendix~A.2 of \citealt{ahlborn2020} \citep[see also ][]{klion2017}. Using these stellar models and rotation profiles, we compute synthetic observations from the base of the RGB up to the bump in luminosity. Each synthetic observable set contains 12 dipole modes (4 radial orders, chosen symmetrically around the stellar model value of $\nu_{\rm max}$) with associated rotational splittings computed from Eq.~(\ref{eqsplittings}) and synthetic uncertainties \citep[see ][Appendices~A.3 and A.4]{ahlborn2020}. 
	
	As described in Sec.~\ref{secMOLA}, we introduced a second trade-off parameter $\theta$ for the eMOLA inversions to balance the two parts of the objective function. We set the first trade-off parameter $\mu$ to a fixed value of 1 for the core and 1000 for the surface to reduce the observational uncertainties, and then calibrate $\theta$ (see Sec.~\ref{sectheta} and Appendix~\ref{secmu} for details).  Using the calibrated trade-off parameters, we construct a core ($r_0/R=0.003$) and a surface ($r_0/R=0.98$) AK using eMOLA inversions for a 1.0~$\text{M}_\odot$, solar metallicity model at the base of the RGB with $L=7.67~\text{L}_\odot$ and $T_\text{eff}=4690~\text{K}$. In the remainder of this section and in Sec.~\ref{seclinsplit} and \ref{secdiss} we use this model as our test model. The results for the MOLA inversions are taken from \cite{ahlborn2020}.
	\subsection{Surface rotation}
\begin{figure}
		\centering
		\includegraphics[scale=1]{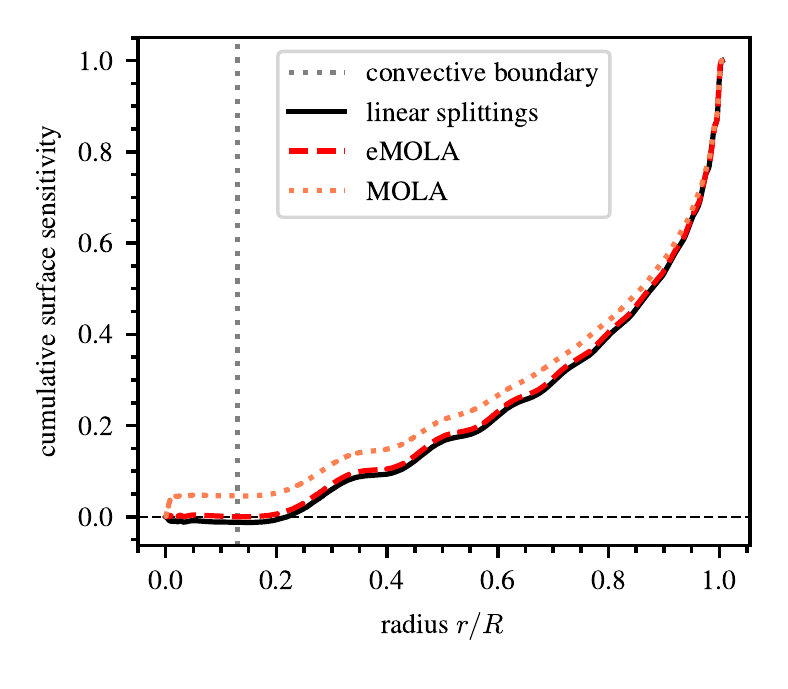}
		\caption{Comparison of cumulative sensitivity kernels for the surface $\Omega_\mathrm{surf}$ rotation rate as a function of radius. The results obtained with the extended MOLA and the MOLA inversions are shown with red and light-red lines, respectively. The grey dotted line indicates the base of the convective envelope. The black line shows the cumulative sensitivity kernel for the surface rotation obtained with the LSA method. The construction of the sensitivity kernels applying the LSA is described in Sec.~\ref{seclinsplit}. The test model described at the beginning of Sec.~\ref{seceMOLAsingle} has been used for the computation of the synthetic data.}
		\label{figcumuleMOLA}
	\end{figure}

	The cumulative SAK obtained from the eMOLA inversions is shown in Fig.~\ref{figcumuleMOLA}. It does not show any cumulative sensitivity in the deep core of the star, and only starts to accumulate sensitivity beyond the base of the convective envelope. Such a SAK allows us to obtain an estimate of the average envelope rotation, which is independent of the core rotation. This is in contrast to the results of MOLA inversions for red-giant stars on the lower part of the RGB \citep{ahlborn2020}. The cumulative SAK obtained from MOLA inversions for the same stellar model and synthetic data is also shown in Fig.~\ref{figcumuleMOLA}. Contrary to the cumulative SAK from the eMOLA inversions, the MOLA AK shows noticeable cumulative sensitivity to the core. In this model, it amounts to about 5\%. According to Eq.~(\ref{eqomegaavg}) the rotation rate is estimated as the integral over the rotation profile with the AK acting as a weight function. Due to the fast core rotation, the cumulative sensitivity of the MOLA SAK to the core translates to a systematic error of the estimated surface rotation rate of about 30\% (30~nHz for the present case) when assuming the envelope step profile. 
	
	We show the estimated rotation rates in Table~\ref{tablerateseMOLA} for the three different synthetic rotation profiles. When assuming the envelope or core step rotation profiles, the surface rotation is recovered within 1$\sigma$ uncertainties. In the case of the convective power law profile, the estimated surface rotation gets closer to the input than the MOLA result, although it still deviates substantially. This arises because in the synthetic profile, the rotation rate is not constant in the envelope, as it increases with depth in regions to which the SAK is still sensitive. That being said, eMOLA also decreases the deviation for this profile when compared to the MOLA inversions. 
	\begin{table}
		\caption{Comparison of core and surface rotation rates obtained with eMOLA inversions}
		\label{tablerateseMOLA}
		\centering
		\begin{tabular}{c c c c}
			\hline\hline
			\rule{0pt}{12pt}&&Core &Surface\\
			Profile&Method&rotation&rotation\\
			&&[nHz]&  [nHz]\\
			\hline
			\rule{0pt}{10pt}Core step &eMOLA&$690\pm6$ &$102\pm6$\\
			&\hphantom{e}MOLA&$686\pm6$&$128\pm7$\\
			\rule{0pt}{10pt}Envelope step&eMOLA & $749\pm6$ &$100\pm6$\\
			&\hphantom{e}MOLA&$745\pm6$&$129\pm7$\\
			\rule{0pt}{10pt}Convective power law&eMOLA& $750\pm6$ &$152\pm6$\\
			&\hphantom{e}MOLA&$747\pm6$&$179\pm7$\\
			\hline
			\rule{0pt}{10pt}Input& & 750 &100\\
			\hline  
		\end{tabular}
		\tablefoot{The profile type used to compute the synthetic rotational splittings is given in the first column. Core and surface rotation rates have been calculated with target radii of $r_0/R=0.003$ and $r_0/R=0.98$ respectively. The last row contains the input values for core and surface rotation rates in the different profiles.}
	\end{table}
	\subsection{Core rotation}
\begin{figure}
		\centering
		\includegraphics[scale=1]{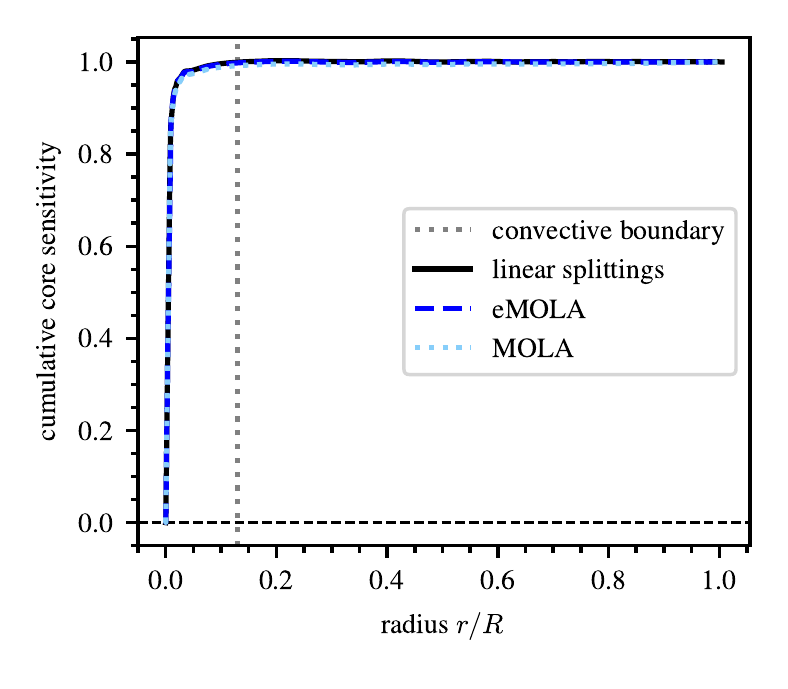}
		\caption{Comparison of cumulative sensitivity kernels for the core $\Omega_\mathrm{core}$ rotation rate as a function of radius for synthetic data. The results obtained with the extended MOLA and the MOLA inversions are shown with blue and light-blue lines, respectively. The grey dotted line indicates the base of the convective envelope. The black line shows the cumulative sensitivity kernel for the core rotation obtained with the linear splittings approximation. The construction of the sensitivity kernels applying the LSA is described in Sec.~\ref{seclinsplit}.}
		\label{figcumul}
	\end{figure}

	The cumulative core averaging kernel (CAK) obtained from the eMOLA inversion is shown in Fig.~\ref{figcumul}. The cumulative CAK rises steeply in the deep interior and very quickly reaches a value close to one. This indicates that it is very well localised in the core. For comparison, the result from a MOLA inversion for the same stellar model and synthetic data set as used previously is also shown in Fig.~\ref{figcumul}. The cumulative CAK obtained with the eMOLA and MOLA inversions show very similar behaviour. This allows us to compute a well-localised average value of the core rotation with either of the two inversion methods. The core rotation rates estimated with eMOLA are given in the first column of Table~\ref{tablerateseMOLA}. For the envelope step and the convective power law profiles, the core rotation rate is recovered within the 1$\sigma$ uncertainties. For the core step profile, the estimated rate deviates substantially from the input value. Even though the CAK is well localised, there is still substantial sensitivity to the slow rotating envelope beyond $1.5r_\text{H}$ which decreases the estimate for the core rotation rate. This is in very close agreement with results from MOLA inversions.
	\subsection{Calibration of the trade-off parameter $\theta$}
	\label{sectheta}
	As discussed above, the introduction of a second trade-off parameter is necessary to balance the original MOLA term and the newly introduced extension term in the eMOLA objective function. This parameter needs to be calibrated for each model individually. In principle, $\theta$ could be calibrated for both the core and the SAK independently. Previous studies have shown that SAKs are more prone to a leakage of the sensitivity out of the region of interest. Hence, we only calibrate the parameter for the surface, and check whether the CAK is also localised for the same parameter value. As a criterion for an optimal trade-off parameter value $\theta$, we have chosen the cumulative sensitivity below the base of the convection zone. The trade-off parameter $\mu$ is fixed while calibrating $\theta$. For a brief discussion of how to choose a suitable value of $\mu$ we refer to Appendix~\ref{secmu}.
	
	\begin{figure}
		\centering
		\includegraphics{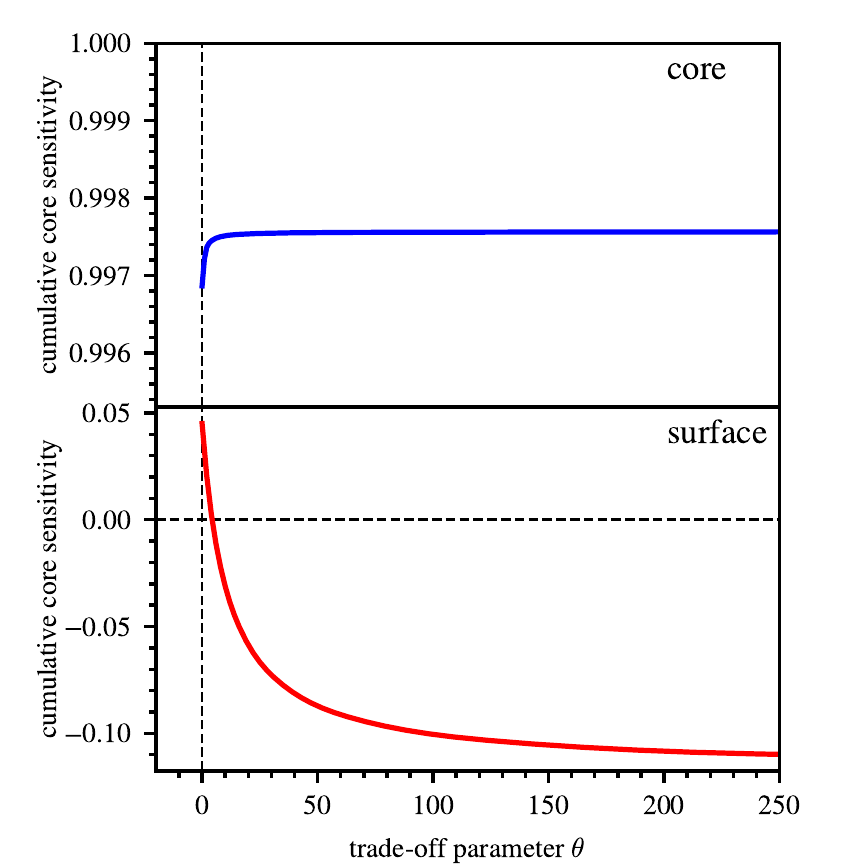}
		\caption{Cumulative sensitivity below the base of the convection zone for a core and a surface averaging kernel for synthetic data in the upper and lower panels, respectively. The trade-off parameter $\mu$ has been set to 0.}
		\label{figcalib}
	\end{figure}
	
	The cumulative core sensitivity as a function of the trade-off parameter $\theta$ is shown for a core and a surface AK in Fig.~\ref{figcalib} for the same model as discussed in the previous subsections. For a value of ${\theta=0}$, the eMOLA inversion is identical to the MOLA inversion and the SAK has about 5\% sensitivity to the core rotation. For an increasing trade-off parameter $\theta$, the impact of the extension term becomes more and more important. Increasing $\theta$ causes a monotonic decrease in the cumulative sensitivity of the SAK to the core rotation. We choose the optimal value of the trade-off parameter $\theta$ at the zero-crossing of the cumulative core sensitivity. When increasing the trade-off parameter further, the SAK starts to show negative side-lobes in the core region and to accumulate negative sensitivity to the core rotation. Asymptotically for $\theta \to \infty$ the eMOLA inversion behaves as if the objective function would consist of only the extension term. Therefore the cumulative core sensitivity converges to a constant value, as shown in Fig.~\ref{figcalib} for high $\theta$. The transition from positive to negative cumulative core sensitivity indicates that a SAK constructed with either of the two terms of the objective function Eq.~(\ref{eqobj2}) alone would not have the desired properties. To obtain a SAK which has a small amplitude and no cumulative sensitivity away from the target radius simultaneously, both terms in the objective function are needed. 
	
	The calibration procedure works for models that are far enough evolved into the red-giant phase, where the base of the convection zone is located relatively close to the core and below the lower boundary of the p-mode cavity. In younger red-giant models, in which the p-mode cavity extends below the base of the convective envelope, the parameter should be calibrated differently for eMOLA to reduce the core sensitivity. However, in these less evolved red-giant stars, the oscillation modes are less sensitive to the deep core layers, such that the application of MOLA is sufficient to obtain localised SAKs, and $\theta$ can be set to 0 (see Fig.~\ref{figsenseMOLA} and~\ref{figtheta} below). In the following, we will focus on more evolved models in which the application of eMOLA inversions leads to improved estimates of the surface rotation rate.
	
	The upper panel of Fig.~\ref{figcalib} shows the cumulative core sensitivity of the CAK as a function of $\theta$. Generally, CAKs are more localised than SAKs in red giants. Therefore, already for $\theta=0$, more than 99\% of the sensitivity is confined below the base of the convection zone. For increasing trade-off parameter, the sensitivity to the core rotation slightly increases, which means that the CAK becomes more localised. For high values of $\theta$, again a converged state is reached. This behaviour is very favourable for our analysis. For any positive trade-off parameter we calibrate using the SAK, we will obtain a well-localised CAK. Therefore, we conclude that the usage of a single value for the trade-off parameter $\theta$ for both the core and the SAK is sufficient.
	\section{Inversions along the lower part of the red-giant branch}
	\label{secinvRGB}
	After studying the results of eMOLA inversions for a particular stellar model at the base of the RGB, we now investigate how the inversion results evolve along the lower part of the RGB, as this is the range in which rotational splittings are expected to be observed. The synthetic data used are described in the beginning of Sec.~\ref{seceMOLAsingle}. As the measurement of rotation rates at intermediate radii is difficult with low degree modes, we limit this study to dividing the star into a core and a surface zone. The sensitivity of the AK in the respective zone is computed with the following two integrals:
	\begin{align}
	\beta_\text{core}(r_0)&=\int_0^{r_\text{core}}K(r,r_0)\,\text{d}r,\\
	\beta_\text{surf}(r_0)&=1-\beta_\text{core}(r_0).
	\end{align}
	where we chose $r_{\rm core}=r_{\rm rcb}$ to separate the inner region of the star from the convective envelope.
	\begin{figure}
		\centering
		\includegraphics[scale=1]{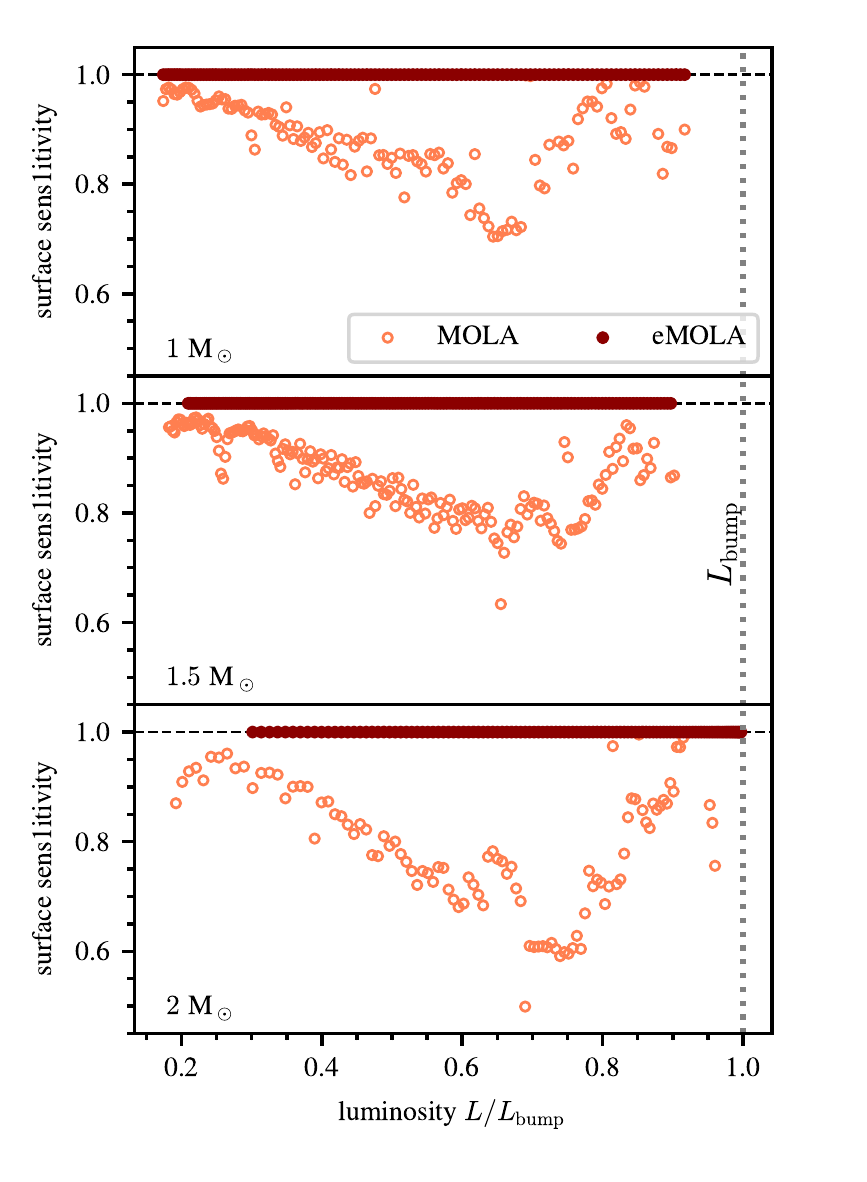}
		\caption{Comparison of surface sensitivities $\beta_\text{surf}=\beta_\text{surf}(0.98)$ obtained with MOLA and eMOLA inversions as a function of luminosity for models along the 1.0, 1.5 and 2.0~$M_\odot$ evolutionary tracks. A sensitivity of 1 indicates that the surface averaging kernel probes only the convective envelope, whereas a sensitivity smaller than 1 indicates that the estimate of the convective envelope rotation rate is contaminated by the core rotation rate.}
		\label{figsenseMOLA}
	\end{figure}	
	We now define the core sensitivity $\beta_\text{core}=\beta_\text{core}(0.003)$ which is the sensitivity of the CAK in the core; and the surface sensitivity $\beta_\text{surf}=\beta_\text{surf}(0.98)$ which is the sensitivity of the SAK in the envelope. The surface sensitivities obtained with MOLA and eMOLA are shown in Fig.~\ref{figsenseMOLA} for three different masses as a function of the stellar luminosity. The surface sensitivities obtained with eMOLA inversions are essentially constant with a value of one all along the RGB. This shows that the calibration of the parameter $\theta$ efficiently reduces the cumulative sensitivity below the base of the convective envelope. We find this result to be independent of the mass in the ranges investigated. This means that for eMOLA inversions, the sensitivity of the SAK is exclusively confined to the convective envelope. As for the individual model studied above, the estimated surface rotation rate is an average value of the whole convective envelope. As the models evolve up the RGB, the MOLA SAKs accumulate substantial sensitivity to the core rotation, which is reflected by the dip in the surface sensitivities of the MOLA inversions in Fig.~\ref{figsenseMOLA}. 
	
	\begin{figure}
		\centering
		\includegraphics{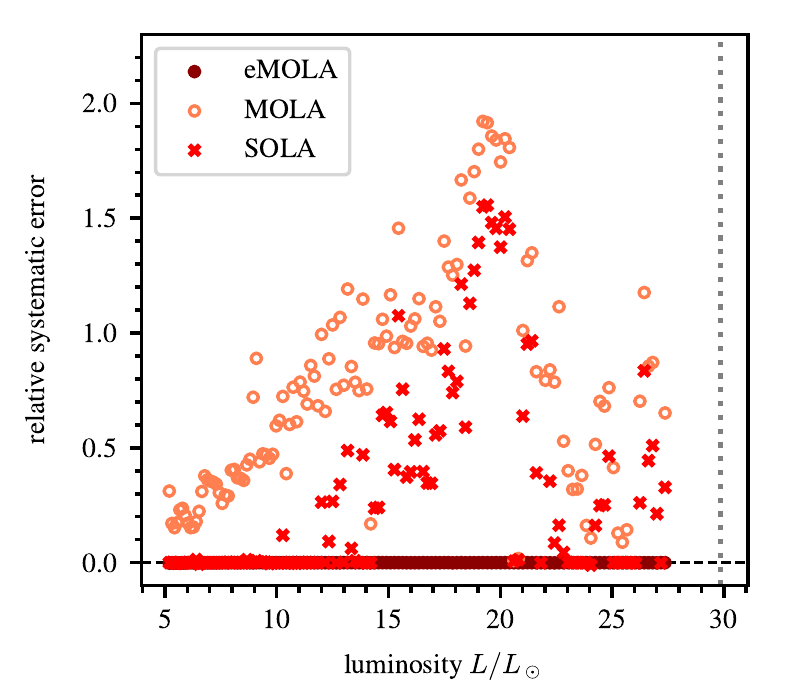}
		\caption{Relative systematic errors of the envelope rotation rates as a function of luminosity for MOLA, SOLA and eMOLA inversions shown with light red, red and dark red points respectively for the 1~$\text{M}_\odot$ track. The bump luminosity is shown with a vertical grey dotted line. Here, the envelope step profile has been used. Results for the 1.5 and 2~$\text{M}_\odot$ tracks show the same behaviour. In contrast to MOLA and SOLA inversions, the eMOLA results do not suffer from systematic errors.}
		\label{figsyserr}
	\end{figure}
	In addition to the surface sensitivities, we show in Fig.~\ref{figsyserr} the relative systematic errors of the surface rotation rates as a function of luminosity. In addition to the MOLA and eMOLA results, we also show results obtained from subtractive optimally localised averages (SOLA) inversions. A detailed discussion of the SOLA method and the calibration of the inversion parameters can be found in Appendix~\ref{secSOLA}. Following previous works, we have chosen a Gaussian target function. Note that other choices for the target function are possible depending on the specific application. In contrast to the MOLA and SOLA results, surface rotation rates from eMOLA can be recovered with very high accuracy. As inferred from Fig.~\ref{figsenseMOLA} the SAKs do not show cumulative sensitivity below the base of the convective envelope. As a consequence, the systematic errors will vanish when assuming the envelope step profile, which is constant below and above the base of the convective envelope. For the original MOLA inversion, these systematic errors rise up to 200\% (see light red points in Fig.~\ref{figsyserr}). The systematic errors obtained from the SOLA inversions show a bimodal behaviour. For less evolved stars, we were able to suppress the core sensitivity. The systematic errors are hence reduced to zero. As of a certain evolutionary state, the systematic errors from SOLA rise however steeply to a similar level as seen from the MOLA inversions. As for the MOLA inversions, this rise in systematic errors can be also explained by the increase of the core sensitivity of the oscillation modes. In contrast to the MOLA inversions, SOLA inversions can suppress the core sensitivity in less evolved red giants by tuning the width of the target function $\Delta$ (see Appendix~\ref{secSOLA} for details). This is similar to the eMOLA inversions in which we suppress the core sensitivity by tuning $\theta$. For more evolved stars it was however no longer possible to find the SOLA inversion parameter such that the core sensitivity is completely suppressed. We attribute this to the choice of the SOLA objective function. For the details of this effect, we refer to Sec.~\ref{secSOLA} in the appendix. This is in contrast to the eMOLA inversions that allow us to suppress the core sensitivity along the whole evolutionary track. Similarly to the envelope step profile that we show here, for the core step and convective power law profiles, the systematic errors obtained from the eMOLA inversions are significantly reduced compared to MOLA inversions (see Appendix~\ref{secsyserr} for a short discussion). This shows that the suppression of the cumulative core sensitivity in the SAK allows us to probe the average envelope rotation rate very accurately along the lower part of the RGB. From the base of the RGB up to the bump in luminosity, this will enable additional constraints to be placed on the angular momentum transport, provided that rotational splittings can be measured in the power spectra of such stars. We would like to point out that this is achieved with a set of 12 dipole modes only.
	\begin{figure}
		\centering
		\includegraphics[scale=1]{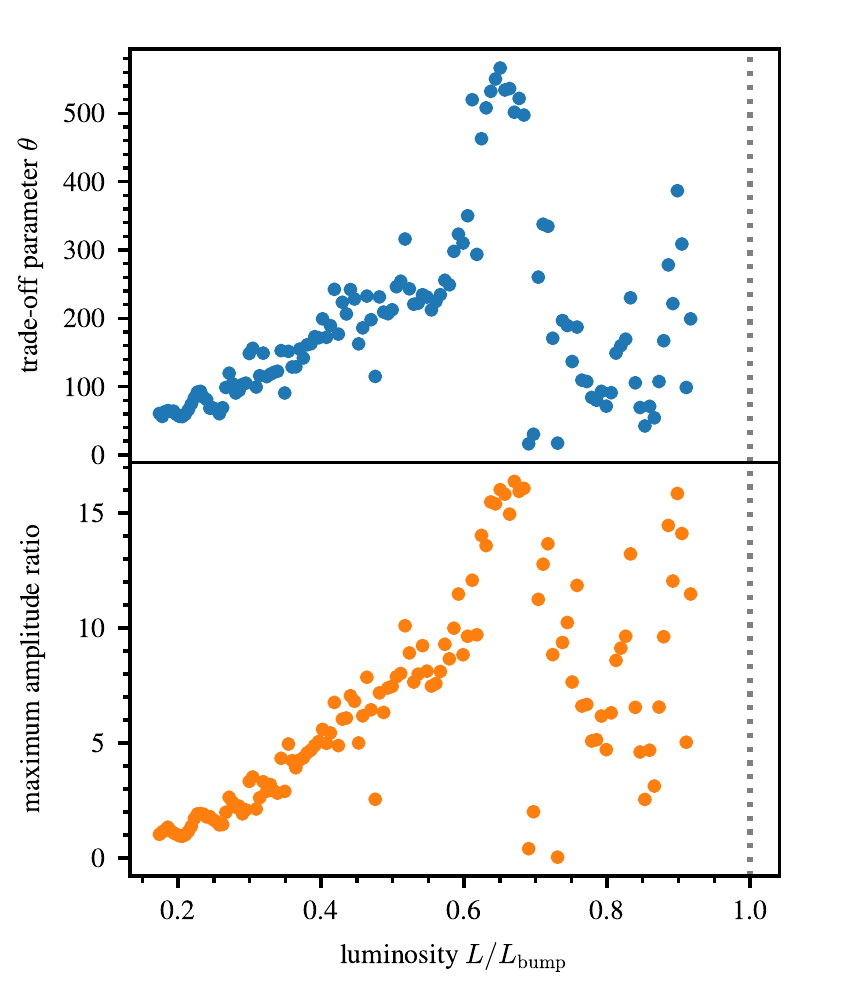}
		\caption{Calibrated trade-off parameter $\theta$ and maximum amplitude ratio as a function of luminosity in the upper and lower panels, respectively, for the $1\text{M}_\odot$ track of solar metallicity. The maximum amplitude ratio is computed as the ratio of the maximum amplitude of the surface averaging kernel in the core to the maximum amplitude of this kernel in the envelope. The grey dotted line indicates the bump luminosity.}
		\label{figtheta}
	\end{figure}
	\cite{ahlborn2020} concluded that the dipolar oscillation modes evolve to become more and more sensitive to the core due to the peak in the buoyancy frequency caused by the discontinuity in mean molecular weight. As shown in Fig.~\ref{figsenseMOLA}, we do not find an increased cumulative sensitivity to the core rotation any more when applying eMOLA inversions. By construction, the trade-off parameter $\theta$ is calibrated such that the cumulative sensitivity to the core is suppressed. The upper panel of Fig.~\ref{figtheta} shows the value of the calibrated trade-off parameter $\theta$ as a function of stellar luminosity. More evolved stars need a greater value of $\theta$ due to the increased core sensitivity of their individual rotational kernels. This effect is visible in Fig.~\ref{figtheta} where $\theta$ clearly increases towards higher luminosities and shows a maximum around ${0.6-0.8\,L_\text{bump}}$, which coincides with the minimum in surface sensitivity observed in \cite{ahlborn2020} (compare to the MOLA results in Fig.~\ref{figsenseMOLA}). For younger stars, toward the base of the RGB, $\theta$ tends toward zero. As in young red giants or subgiants the oscillation modes are less sensitive to the core, less core sensitivity needs to be suppressed. At some point, a parameter value of ${\theta=0}$ is sufficient to obtain the envelope rotation, which is equivalent to applying MOLA.
	
	Another way of looking at this is to analyse the behaviour of the maximum amplitudes in the core and in the envelope. In the lower panel of Fig.~\ref{figtheta} we show the ratio of the maximum amplitude of the SAK in the core and in the envelope as a function of luminosity. The resulting shape looks very similar to that of the trade-off parameter shown in the upper panel. When evolving up the RGB, the core amplitude of the SAK increases until it reaches a maximum again at around ${0.6-0.8\,L_\text{bump}}$. This increase in the amplitude does not however translate into an increased sensitivity as shown above. The small scale oscillations in the core component of the SAKs cancel each other to result in approximately zero cumulative sensitivity to the core region. The results for the trade-off parameter and the maximum amplitude together indicate that in fact the stellar oscillation modes become more sensitive to the core when evolving up the RGB. The main difference between this and the previous works is that eMOLA inversions can avoid this increased sensitivity to accumulate in the core, contrary to MOLA inversions.

	\section{Linearised rotational splittings}
	\label{seclinsplit}
	\cite{goupil2013} showed that the rotational splittings of red-giant stars depend approximately linearly on the mode trapping $\zeta$. We refer to this as the linear splittings approximation (LSA) method. The mode trapping is defined as the ratio of mode inertia in the core $I_{\rm core}$ to the total mode inertia $I$:
	\begin{align}
	\zeta=\frac{\int_0^{r_\mathrm{core}}(\xi_r^2+l(l+1)\xi_h^2)\rho r^2\mathrm{d}r}{\int_0^R(\xi_r^2+l(l+1)\xi_h^2)\rho r^2\mathrm{d}r}=\frac{I_{\rm core}}{I},
	\end{align}
	where $\xi_r$ and $\xi_h$ are the radial and horizontal displacement eigenfunctions and $\rho$ is the density profile of the stellar model. In this section, we compare the eMOLA inversion results to the LSA method and show their near equivalence.
	\subsection{Linear regression coefficients}
	\label{seclinsplit2}
	The linear dependence of the rotational splittings on the mode trapping can be conveniently written in matrix notation:
	\begin{align}
	\vec{\delta\omega}=\mathbf{Z}\cdot\vec{\beta}+\vec{\sigma},	\label{eqlin}
	\end{align}
	where $\vec{\delta\omega}=(\delta\omega_1,...,\delta\omega_M)^T$, $\mathbf{Z}=(\vec{1}, \vec{\zeta})$, $\vec{\sigma}=(\sigma_{\delta\omega_1},...,\sigma_{\delta\omega_M})^T$ and $\vec{\beta}=(\beta_0,\beta_1)^T$. Here, $\beta_0$ and $\beta_1$ are parameters of the linear model. It can be shown that the parameters of the linear relation Eq.~(\ref{eqlin}) depend on the core and the envelope rotation rates:
	\begin{align}
	\Omega_\mathrm{core}&=2(\beta_0+\beta_1)\label{eqrate1},\\
	\Omega_\mathrm{surf}&=\beta_0\label{eqrate2}.
	\end{align}
	We have determined $\vec{\beta}$ by means of linear regression. The parameters are estimated as:
	\begin{align}
	\vec{\beta}=\underbrace{(\mathbf{Z}^T\cdot \mathbf{W}\cdot \mathbf{Z})^{-1}\mathbf{Z}^T \mathbf{W} }_{\mathbf{C}}\cdot\,\vec{\delta\omega}.\label{eqparam}
	\end{align}
	Here, the usage of a weight matrix $\mathbf{W}$ is required as the uncertainties of the rotational splittings are non-uniform across the frequency range. We choose $\mathbf{W}$ to be a diagonal matrix with ${W_{ii}=1/\sigma_{\delta\omega_i}^2}$.
	
	The matrix $\mathbf{C}$ can be interpreted as a coefficient matrix determining the impact of each data point on the derived parameters~$\vec{\beta}$. The first row of $\mathbf{C}$ contains the coefficients for $\beta_0$, whereas the second row contains the coefficients for $\beta_1$. Comparable to the computation of the AKs in MOLA inversions as given by Eq.~(\ref{eqavgkernel}) we use the coefficients $c_{\beta_j}$ corresponding to $\beta_j$ to construct sensitivity kernels (SKs) for the slope and the intercept:
	\begin{align}
	K_{\beta_j}(r)=\sum_{i\in\mathcal{M}}c_{\beta_j,i}\mathcal{K}_i(r).\label{eqkernel}
	\end{align}
	This is analogous to the AK analysis of the asteroseismic frequency ratios by \cite{oti2005}. Following Eqs.~(\ref{eqrate1}) and~(\ref{eqrate2}) we construct SKs for the core and surface rotation rates as determined from the linear approximation:
	\begin{align}
	K_\mathrm{core}(r)&=2[K_{\beta_0}(r)+K_{\beta_1}(r)],\label{eqkernelcore}\\
	K_\mathrm{surf}(r)&=K_{\beta_0}(r).\label{eqkernelsurf}
	\end{align}
	We will use these SKs to interpret the results obtained from the linear approximation.
	\subsection{Sensitivity kernels}
	\label{secsenskernels}
\begin{figure}
		\centering
		\includegraphics[scale=1]{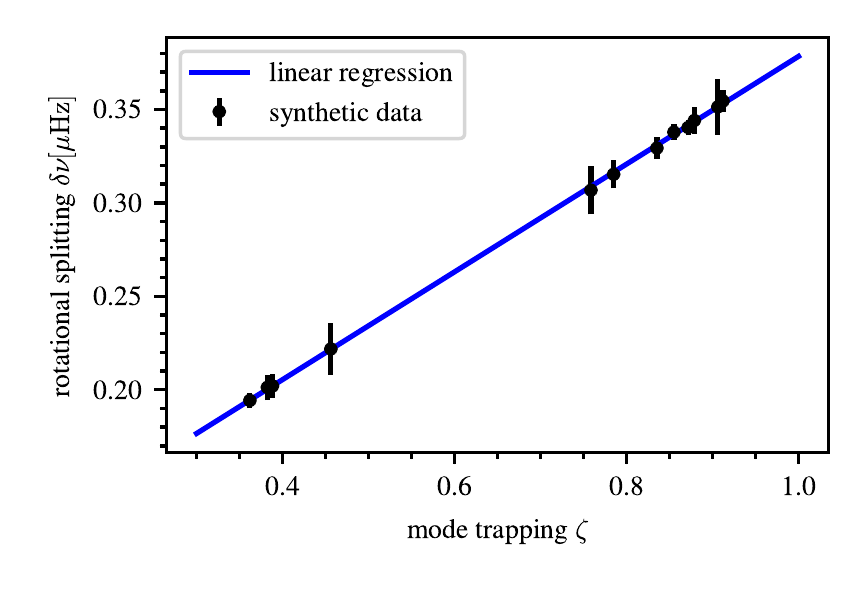}
		\caption{The rotational splittings $\delta\omega$ as a function of the mode trapping $\zeta=I_{\rm core}/I$ are shown with the black marker. The blue line indicates a fit to the data obtained by means of linear regression as described by Eq.~(\ref{eqparam}). Note that modes with a low $\zeta$ are p-dominated, while modes with a high $\zeta$ are g-dominated.}
		\label{figdatafit}
	\end{figure}

	In this section, we apply the analysis described in Sec.~\ref{seclinsplit2} to a set of synthetic data. The computation of the synthetic data and the red-giant test model used here are the same as described in Sec.~\ref{seceMOLAsingle}. Figure~\ref{figdatafit} shows the rotational splittings as a function of the mode trapping for the red-giant model. We chose the envelope step rotation profile to compute the rotational splittings. The synthetic rotational splittings follow the linear relation very closely. This shows that the linear approximation works very well for this synthetic rotation profile. The good agreement can be understood because both the linear approximation as well as the envelope step rotation profile describe the star as being composed of only two zones.
	Using this set of rotational splittings and mode trappings, we compute the core and surface rotation rates as defined by Eqs.~(\ref{eqrate1}) and~(\ref{eqrate2}). The results for the three different synthetic rotation profiles are shown in Table~\ref{tablerateslinsplit}. The core rotation rates are recovered within the 2$\sigma$ uncertainties for the envelope step and the convective power law rotation profiles. For the core step profile, a substantial deviation from the input rotation rate is observed. This is consistent with results obtained using MOLA inversions. For the surface rotation rate, the results also agree within $2\sigma$ with the input rotation rate for the core step and the envelope step rotation profile. For the convective power law profile, the deviations exceed the formal 2$\sigma$ uncertainties. These LSA results are improved compared to the rotation rates as computed using MOLA inversions, and the eMOLA inversions provide even more accurate rotation rates.
	\begin{table}
		\caption{Comparison of core and surface rotation rates obtained with the linear rotational splittings approximation according to Eqs.~(\ref{eqrate1}) and~(\ref{eqrate2})}
		\label{tablerateslinsplit}
		\centering
		\begin{tabular}{c c c}
			\hline\hline
			\rule{0pt}{12pt}&Core &Surface\\
			Profile&rotation&rotation\\
			&[nHz]&  [nHz]\\
			\hline
			\rule{0pt}{10pt}core step &$697.3\pm5.8$ &$92.5\pm5.6$\\
			\rule{0pt}{10pt}envelope step & $756.8\pm5.8$ &$90.1\pm5.6$\\
			\rule{0pt}{10pt}convective power law & $758.0\pm5.8$ &$141.5\pm5.6$\\
			\hline
			\rule{0pt}{10pt}input & 750 &100\\
			\hline                  
		\end{tabular}
		\tablefoot{For reference, the input values of the synthetic rotation profiles in the core and at the surface are given in the bottom row. The profile type used to compute the synthetic rotational splittings is given in the first column.}
	\end{table}
	
\begin{figure}
		\centering
		\includegraphics[scale=1]{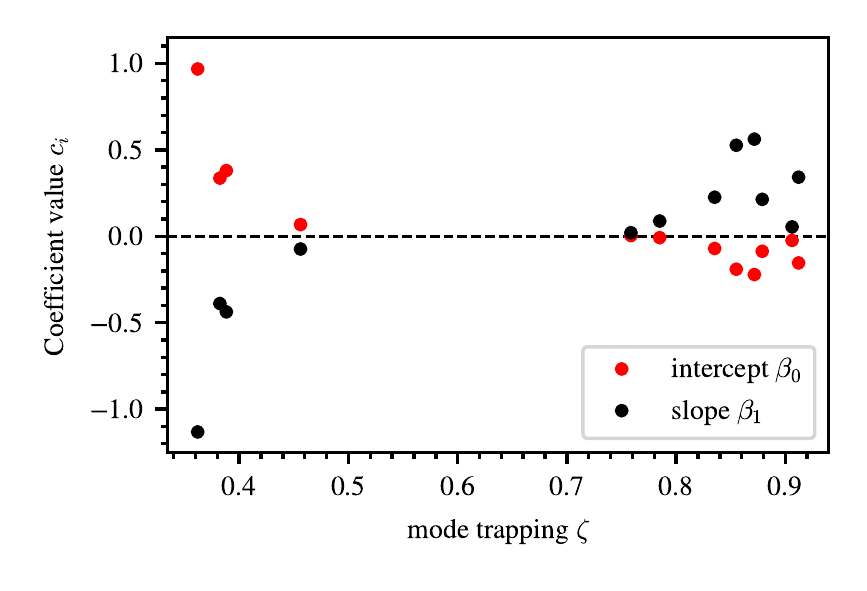}
		\caption{Linear regression coefficients as a function of mode trapping $\zeta=I_{\rm core}/I$. The coefficients for the slope $\beta_1$ are shown in black whereas the coefficients for the intercept $\beta_0$ are shown in red.}
		\label{figcoeff}
	\end{figure}

	To construct SKs for the obtained core and surface rotation rates, we proceed to compute the coefficients of the linear regression as described by the matrix $\mathbf{C}$ in Eq.~(\ref{eqparam}). The coefficients for the slope ($\beta_1$) and the intercept ($\beta_0$) are shown in Fig.~\ref{figcoeff} as a function of the mode trapping. Although the distribution of the sensitivity can only be interpreted when the coefficients are multiplied with the SKs of the rotational splittings, the coefficients allow already for some preliminary conclusions. The coefficients for the intercept have larger values for lower mode trappings, while for higher trappings the coefficient values are smaller in magnitude and negative. According to the definition of the mode trapping, modes with lower mode trappings have more sensitivity in the p-mode cavity. This indicates that the SK of the intercept will be more sensitive to the surface. On the other hand, the coefficients for the slope show values of comparable magnitude for both high and low mode trappings. This indicates that the slope also has substantial sensitivity to the core of the star.
	\begin{figure}
		\centering
		\includegraphics[scale=1]{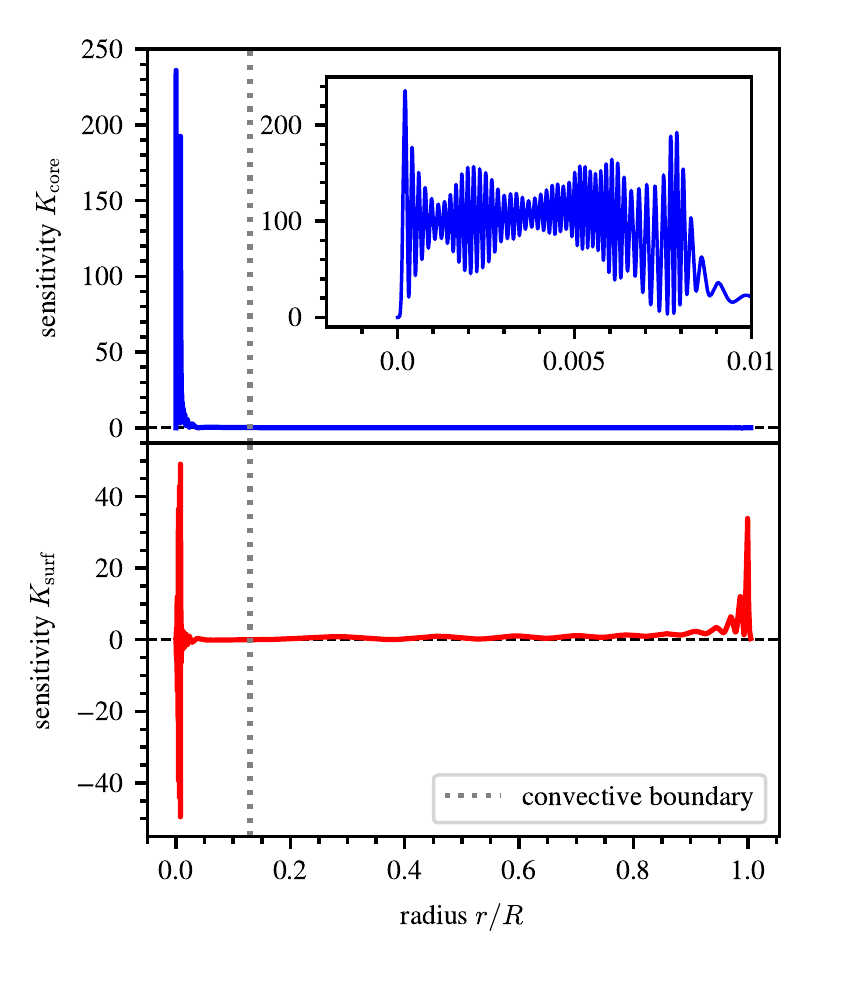}
		\caption{Sensitivity kernels for linear splittings approximation of the core rotation rate $\Omega_\mathrm{core}$ (indicated with the blue line) and the surface rotation rate $\Omega_\mathrm{surf}$ (indicated with the red line) as defined by Eqs.~(\ref{eqrate1}) and~(\ref{eqrate2}) as a function of radius. The grey dotted line indicates the bottom of the convective envelope. The corresponding cumulative sensitivity kernels are shown in Figs.~\ref{figcumuleMOLA} and~\ref{figcumul}.}
		\label{figsens}
	\end{figure}
	
	Using the set of coefficients shown in Fig.~\ref{figcoeff} we compute SKs for the core and surface rotation rate as described by Eqs.~(\ref{eqkernel}), (\ref{eqkernelcore}) and~(\ref{eqkernelsurf}). The resulting SKs are shown in Fig.~\ref{figsens}. The SK of the core rotation shows a clear peak in the deep interior of the star. This indicates that the core rotation rate determined from the linear approximation is exclusively sensitive to the core of the star. We therefore conclude that $\Omega_\mathrm{core}$ as given by Eq.~(\ref{eqrate1}) is an appropriate probe of the mean core rotation. The SK of the surface rotation rate has amplitude throughout the whole stellar interior with a peak in the outer layers of the star. It shows in addition a strong oscillatory behaviour in the core. The oscillatory behaviour originates from the small scale oscillations in the SKs of the individual rotational splittings. The increased sensitivity towards the stellar surface indicates already that the surface rotation rate $\Omega_\mathrm{surf}$ as given by Eq.~(\ref{eqrate2}) is more sensitive to the stellar surface.
	
	To investigate the distribution of the sensitivities, we also compute the cumulative kernels of the two SKs presented above. Fig.~\ref{figcumul} shows the cumulative SK of the core rate. The cumulative kernel of the core rotation rate rises steeply in the very deep interior of the star to a value close to one. This shows that nearly the entire sensitivity of the core rotation rate estimated with Eq.~(\ref{eqrate2}) is confined to the very deep layers. This explains why Eq.~(\ref{eqrate2}) allows us to recover the input rotation rate very accurately. Similarly, Fig.~\ref{figcumuleMOLA} shows the cumulative sensitivity for the surface rotation rate obtained with the linear approximation. It does not show cumulative sensitivity up to a fractional radius of 0.15 in this model. Thereafter, the cumulative kernel shows the build-up of sensitivity throughout the whole convective envelope. This indicates that the surface rotation rate determined from the linear approximation is a measure of the average rotation in the convective envelope. As there is no cumulative sensitivity in the core, the average value is not impacted by the core rotation. As for the core rotation rate, this explains why the linear approximation allows us to recover the input core rotation rate rather accurately. In addition, the SKs, both for the core and the envelope rotation, obtained from the LSA are very similar to the AKs obtained with eMOLA inversions. This further underlines why the LSA results in very accurate estimates for the core and envelope rotation rates. It also shows that the mode set used has in principle the potential to suppress the core sensitivity in the SAK. However, as demonstrated above, MOLA does not suppress the core sensitivity due to the choice of the objective function. With this analysis we have shown a posteriori that the LSA SKs are equivalent to the eMOLA AKs. While for the LSA only a single set of coefficients can be obtained determined by the data and the linear fit, the trade-off parameters $\mu$ and $\theta$ as well as the target-radii allow for influencing the shape of the eMOLA AKs and hence the quality of the estimate of the rotation rates. We therefore conclude that using rotational inversions is still advantageous, despite the simplicity of the LSA method.
	
	\section{Discussion}
	\label{secdiss}
	\subsection{Application of eMOLA inversions}
	\label{secappl}
\begin{figure}
		\centering
		\includegraphics{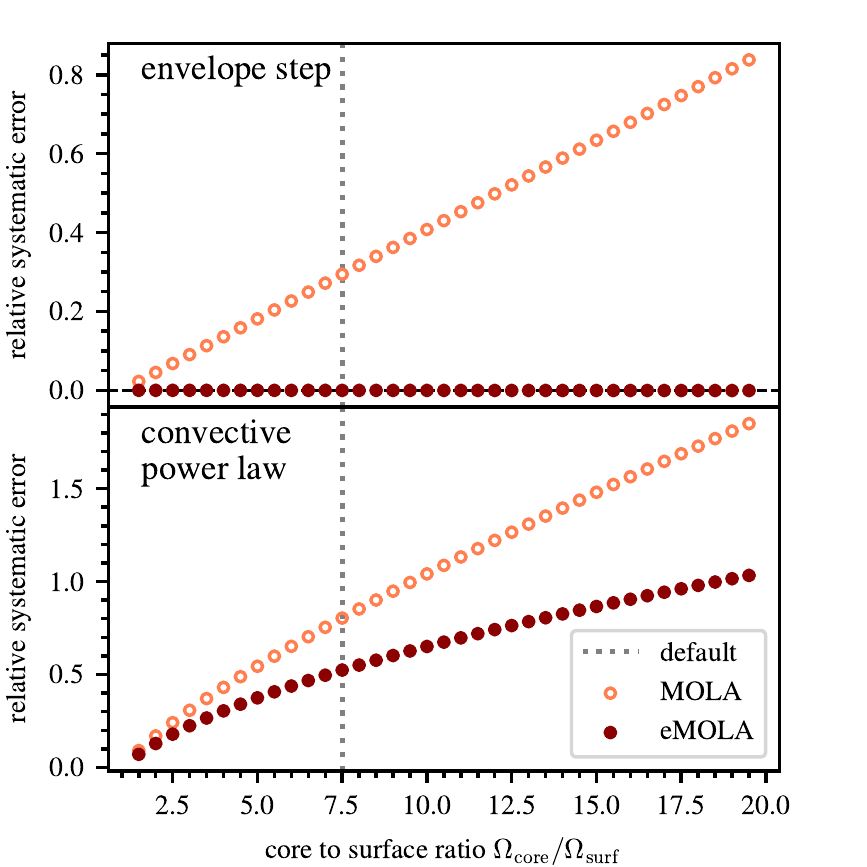}
		\caption{Relative error of the estimated surface rotation rate as a function of the core-to-surface ratio $\Omega_\text{core}/\Omega_\text{surf}$. The upper and lower panels show results for the envelope step and the convective power law profile, respectively. The eMOLA results are shown in dark red and the MOLA results are shown in light red. The grey dotted line refers to the ratio obtained with the core and surface rotation rates of 750~nHz/100~nHz used throughout this paper. Please note that the vertical axis differs in the upper and lower panels. In case of the envelope step profile, the relative errors from eMOLA remain negligibly small for the range of core-to-surface ratios investigated.}
		\label{figrelerrordep}
	\end{figure}

	To demonstrate how eMOLA inversions would work for observed rotational splittings, we compute different sets of synthetic splittings with a range of ratios of the core-to-surface rotation rate for the same test model as described in Sec.~\ref{seceMOLAsingle}. We use the envelope step rotation profile and the convective power law profile to compute the rotational splittings. Subsequently, we use eMOLA inversions to estimate the core and envelope rotation rates. In Fig.~\ref{figrelerrordep}, we show the relative systematic error of the estimated surface rotation rate with respect to the input as a function of the input core-to-surface ratio, ranging from 1.5 up to 20. 
	
	For the envelope step profile, the resulting relative error is interpreted as a linear combination of the core and surface sensitivity multiplied with the rotation rate in the respective cavity. The integral of Eq.~(\ref{eqomegaavg}) can be rewritten as
	
	\begin{align}
	\overline{\Omega}(r_0)=\beta_\text{core}(r_0)\cdot\overline{\Omega_\text{core}}(r_0)+\beta_\text{env}(r_0)\cdot\overline{\Omega_\text{env}}(r_0)
	\end{align}
	by defining the following average core and envelope rotation rates
	\begin{align}
	\overline{\Omega_\text{core}}(r_0)=\frac{\int_0^{r_\text{core}}K(r,r_0)\,\Omega(r)\,\text{d}r}{\beta_\text{core}(r_0)}\\
	\overline{\Omega_\text{env}}(r_0)=\frac{\int_{r_\text{core}}^RK(r,r_0)\,\Omega(r)\,\text{d}r}{\beta_\text{surf}(r_0)}
	\end{align}
	
	This means that for a constant envelope rotation, the estimated surface rotation rate depends linearly on the input core rotation, and the core sensitivity acts as the slope of this relation. The relative errors follow the same linear relation. This linear relation is found when analysing the relative errors of the envelope step profile, shown in the upper panel of Fig.~\ref{figrelerrordep}. For the eMOLA inversions the core sensitivity is suppressed and therefore the slope of the linear relation is essentially non-existent compared to the MOLA inversions. The relative errors remain negligibly small also for a core-to-surface ratio of 20. Therefore, when assuming an envelope step rotation profile, eMOLA inversions recover the envelope rotation rate with very small systematic errors for any reasonable ratio of the core-to-surface rotation. 
	
	For the convective power law profile, the situation is more complicated. To ensure a smooth transition of the rotation profile between core and envelope, the profile in the envelope needs to change as well when changing the core rotation. Contrary to the envelope step profile, the power law contribution in the envelope therefore also increases with increasing core-to-surface ratio. As the SAK shows sensitivity throughout the whole envelope, this change in the envelope rotation will also impact on the relative errors. The relative errors for the convective power law profile are shown in the lower panel of Fig.~\ref{figrelerrordep}. Here, the relative errors do not depend linearly on the core-to-surface ratio. Even though the improvement is less pronounced than for the envelope step rotation profile, the relative error of eMOLA improves by about 40\% compared to MOLA for the core-to-surface rotation rate ratio used throughout this paper. Furthermore, the improvement increases with an increasing core-to-surface ratio.
	\subsection{Dependence on the mode set}
	\begin{figure}
		\centering
		\includegraphics{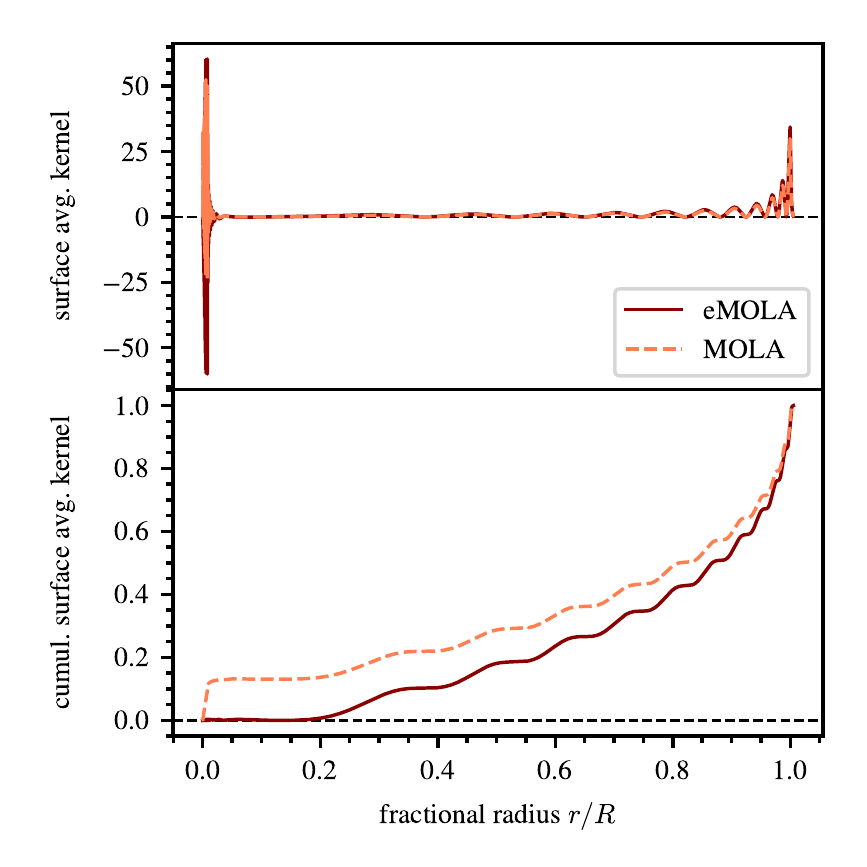}
		\caption{Surface averaging kernels (upper panel) and cumulative surface averaging kernels (lower panel) as a function of the fractional radius computed using different inversion methods for a mode set containing one p- and one g-dominated mode. The results for the eMOLA inversion are shown with a solid dark red line and the MOLA results are indicated with the light red dashed line. In contrast to the eMOLA results, the MOLA surface averaging kernel shows substantial cumulative core sensitivity.}
		\label{figcumulcompare}
	\end{figure}
	In Secs.~\ref{seceMOLAsingle} and~\ref{secinvRGB} we have used a mode set that always consisted of four radial orders. Per radial order, we have selected a central p- and two g-dominated modes, with lower and higher frequencies (for our test model described in Sec.~\ref{seceMOLAsingle} the radial orders of the selected p-dominated modes are $n_{\rm p}=9,\dots,12$). For comparison, we have computed inversions for four different mode sets (see Table~\ref{tabmodesets}). The first mode set allows comparing to another observationally-motivated mode set, while mode sets two and three are more to test the extreme cases. The fourth mode set is selected based on the importance of the individual oscillation modes for the rotational inversion results. The first mode set is constructed to be similar to the observational results of \cite{dimauro2016} who obtained 14 measured rotational splittings across five radial orders. We have used the same stellar model as in Sec.~\ref{seceMOLAsingle} to compute the synthetic data. This model has similar mass, radius, $\Delta\nu$ and $\nu_{\rm max}$ as the model of \cite{dimauro2016}. The modes for the computation of the synthetic data have been selected to obtain a similar pattern of the synthetic rotational splittings as a function of frequency as in \cite{dimauro2016} (their Fig. 1, panel b). We have used the envelope step rotation profile as the synthetic rotation profile. The second mode set consists of two modes only, one p- and one g-dominated mode, in the vicinity of $\nu_{\rm max}$. The third mode set contains the p-dominated modes of five radial orders ($n_{\rm p}=9,\dots,13$).
	\begin{table}
		\caption{Different mode sets used for rotational inversions}
		\label{tabmodesets}
		\centering
		\begin{tabular}{c c c}
			\hline\hline
			\rule{0pt}{12pt}Name &number of modes& acoustic radial\\
			&& orders\\
			\hline
			\rule{0pt}{10pt}Default&12&4 \\
			\hline
			\rule{0pt}{10pt}Di Mauro-like &14&5\\
			\rule{0pt}{10pt}two modes &2&1\\
			\rule{0pt}{10pt}p-dominated &5&5\\
			\rule{0pt}{10pt}SVD &4&4\\
			\hline                  
		\end{tabular}
		\tablefoot{The second column gives the number of modes in each mode set, the third column gives the number of acoustic radial orders from which modes have been selected. With SVD we refer to the singular value decomposition, see text for details.}
	\end{table}
	For the \cite{dimauro2016}-like mode set, we found only minor differences between the AKs when compared to the default mode set. The SAK computed with the \cite{dimauro2016}-like set shows a smaller amplitude of the strongly oscillating g-mode component. This is a result of the larger number of modes, allowing for combinations $c_i$ that not only cancel out the cumulative sensitivity in the core, but also the absolute value of the AK. For the second mode set, the localisation of the SAK compared with the reference mode set was clearly worse. The SAK computed with the eMOLA inversion is shown in the upper panel of Fig.~\ref{figcumulcompare}. The corresponding cumulative SAK is shown in the lower panel of the same figure. In the envelope, i.e. beyond the base of the convection zone, the SAK looks very similar to the rotational kernel of a single p-mode, which is to be expected as only two modes are included in this set. For comparison, we show the SAK computed with the same mode set using MOLA inversions. In the core, the SAK has slightly smaller amplitude compared to the eMOLA kernel, and it is strongly asymmetric. This asymmetry is reflected in the cumulative sensitivity to the core region visible in the lower panel. Clearly, eMOLA is able to suppress this cumulative sensitivity even when only using two modes. When using only p-dominated modes, the computation of localised AKs was impossible. This is to be expected as all modes of this mode set sound a similar region in the stellar interior. Likewise, we are not able to compute localised core and surface averaging kernels for this mode set when using MOLA inversions. We conclude that eMOLA is not strongly sensitive to the mode set used, as long as p- and g-dominated modes are included. Already for the smallest possible mode set we obtain SAKs that look reasonable. As expected, the localisation of the SAK improves considerably when including more modes, i.e. the oscillatory behaviour of the surface component reduces as well as the amplitude of the core component.
	\begin{figure}
		\centering
		\includegraphics{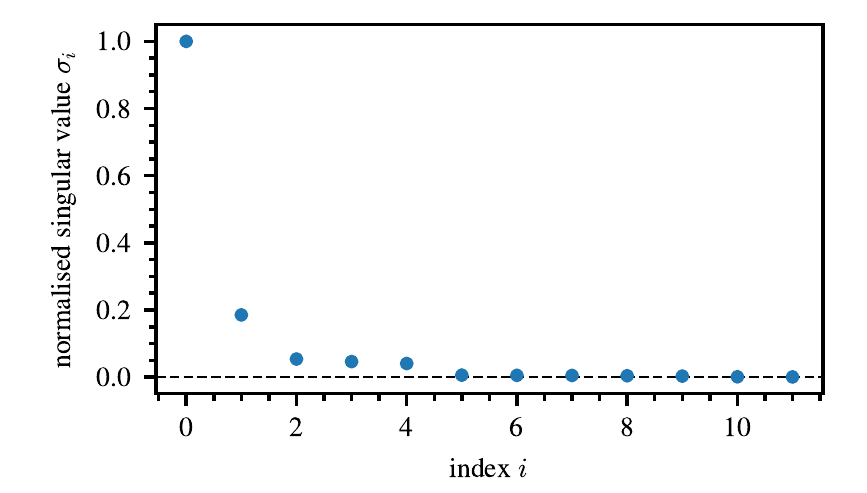}
		\caption{Singular values of the mode set of the test model described in Sec.~\ref{seceMOLAsingle}. The first two singular values clearly exceed the remaining values, indicating that the mode set is dominated by two independent pieces of information.}
		\label{figSVDspectrum}
	\end{figure}
	
	To study the amount of independent information in the rotational kernel data, we have computed the singular value decomposition (SVD) of the default mode set of our test model, as described in Sec.~\ref{seceMOLAsingle}, according to \cite{christensen1993}. The SVD allows us to reformulate the inversion problem in terms of transformed SKs ordered by decreasing importance for the solution. To compute the SVD the continuous inversion problem Eq.~(\ref{eqsplittings}) is rewritten as: 
	\begin{align}
	\vec{\delta\omega}=\mathbf{K}\cdot\vec{\Omega}
	\end{align}
	where $\vec{\delta\omega}$ is the vector of the $M$ rotational splittings as before, $\mathbf{K}$ is a $(M\times N)$ matrix containing the discretisation of the $M$ rotational kernels $\mathcal{K}_i$ on $N$ radial grid points and $\vec{\Omega}$ is a vector containing the discretised rotation profile on $N$ radial grid points. The SVD of the matrix $\mathbf{K}$ is expressed as:
	\begin{align}
	\mathbf{K}=\mathbf{U}\cdot\mathbf{\Sigma}\cdot\mathbf{V}^{\rm T}
	\end{align}
	with orthonormal matrices $\mathbf{U}$ $(M\times M)$ and $\mathbf{V}$ $(N\times N)$ and the $(M\times N)$ diagonal matrix $\mathbf{\Sigma}$ containing the singular values $\sigma_1>\sigma_2>\dots>\sigma_P$ and $P={\rm min}(M,N)$ \citep[e.g.][]{noble2} decreasing in value. 
	\begin{figure}
		\centering
		\includegraphics{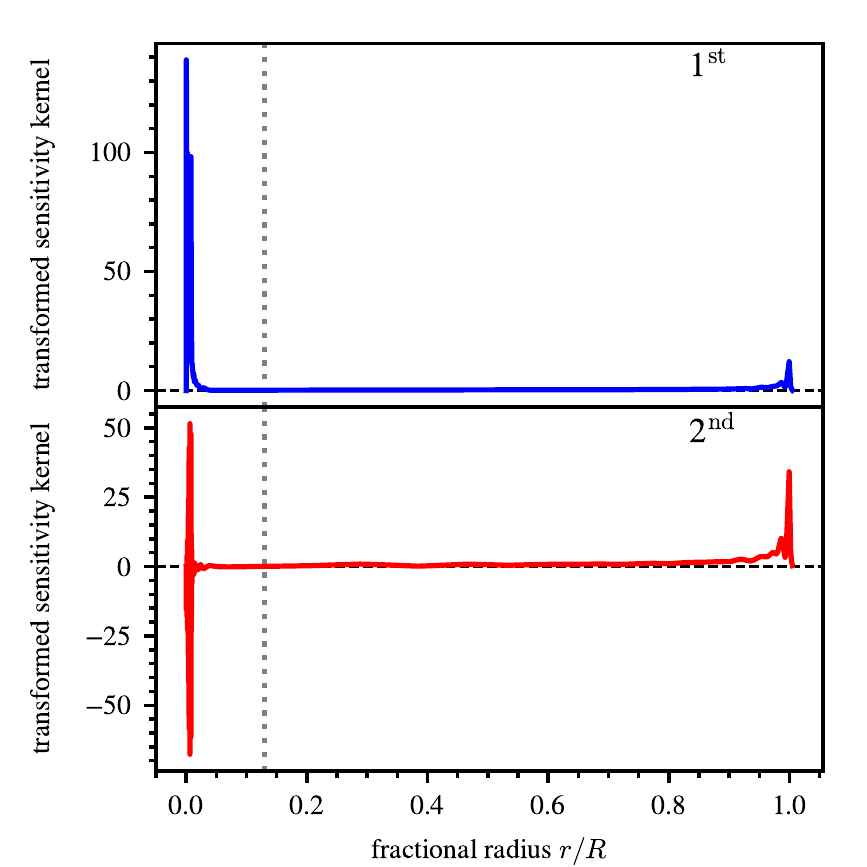}
		\caption{First and second transformed SK of the default mode set, described in Sec.~\ref{seceMOLAsingle} as a function of fractional radius in the upper and lower panel, respectively. The first transformed SK resembles a core averaging kernel, while the second transformed SK resembles a surface averaging kernel. The grey dotted line indicates the base of the convective envelope.}
		\label{figSVDcomponents}
	\end{figure}
	\begin{figure}
		\centering
		\includegraphics{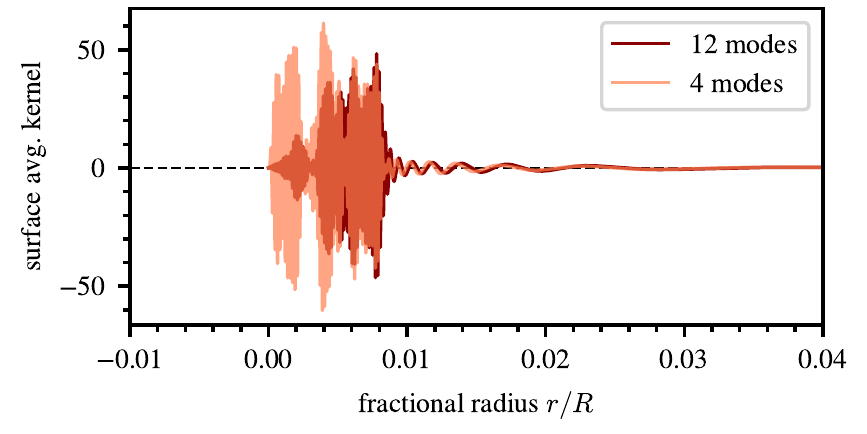}
		\caption{Comparison of the surface averaging kernel in the core region as a function of radius obtained from the eMOLA inversion of the default mode set containing 12 modes and a mode set containing only the four most important modes. Additional modes greatly reduce the SAK's absolute sensitivity to the innermost part of the star.}
		\label{figSVDcomparison}
	\end{figure}
	In Fig.~\ref{figSVDspectrum} we show the normalised singular values $\sigma_i$ ordered by decreasing magnitude as a function of the index $i$. Each singular value is associated with a transformed SK, computed through the coefficients in the matrix $\mathbf{U}$ \citep[see Eq. 11 of][]{christensen1993}. We find that the first two singular values are much larger than the remaining ones, which indicates that the mode set is dominated by the first two transformed SKs. This shows that indeed most oscillation modes sound a similar region in the stellar interior and approximately two independent pieces of information are available to compute the rotational inversions. In Fig.~\ref{figSVDcomponents} we show the transformed SKs associated with the highest and second-highest singular value (shown in Fig.~\ref{figSVDspectrum}) in the upper and lower panel, respectively.  We find that the first transformed SK resembles a CAK, while the second transformed SK resembles a SAK. The analysis of the cumulative integrals of the two transformed SK confirms that the first transformed SK has most of its sensitivity in the deep interior, while the sensitivity of the second transformed SK is more localised towards the surface. In relation with the singular value spectrum shown in Fig.~\ref{figSVDspectrum} this is an intuitive explanation for why it is possible to only probe the rotation in the core and the envelope of this red-giant model.
	
	To construct the fourth mode set, we selected from both the first and the second transformed SKs the two SVD coefficients with the largest magnitude, contained in the matrix $\mathbf{U}$, such that this mode set contains only the four most important modes of our default mode set. We find that a set consisting of only the four most important modes is sufficient to obtain a similar result as for the mode set containing all twelve modes. However, when adding more modes the resulting SAK still improves, i.e. the kernel looks less oscillatory in the envelope and the amplitude of the small scale oscillations in the core is decreased. This can be seen in Fig.~\ref{figSVDcomparison} showing a magnification of the core region of the SAK. When using MOLA inversions for the mode set including the four most important modes, the SAK shows again substantial cumulative sensitivity to the core, as observed for the other mode sets investigated. Finally, we note that the SVD of the mode set consisting of five p-dominated modes possesses only a single dominant singular value, explaining why neither MOLA nor eMOLA are able to computed localised AKs for this mode set.
	
	\subsection{Comparison to other inversion results}
	The MOLA results shown in Fig.~\ref{figsyserr} are well in agreement with other studies using MOLA inversions. In \cite{deheuvels2012} they show a SAK for their early red-giant model in Fig.~12 which shows also about 5\% sensitivity to the core. In \cite{dimauro2016} rotational inversion results obtained with MOLA and SOLA inversions are presented. The latter authors are able to remove the cumulative sensitivity to the core in the SOLA and MOLA SAK. However, no small-scale oscillations are visible in their AKs (see their Figs. 8 and 12). As pointed out above, the presence of small-scale oscillations from the g-mode component are the reason for cumulative core sensitivity in the MOLA SAK and hence their absence may explain the reduced core sensitivity. In \cite{triana2017} a large set of low luminosity red giants is analysed by means of SOLA rotational inversions. In their Fig. 10 and 11 they show SAK and CAK as well as cumulative AKs for KIC~7619745. While the small-scale oscillations of the g-mode component are clearly visible, there is no cumulative sensitivity to the core region in the SAK. As we have discussed above, SOLA inversions are able to suppress the core sensitivity for less evolved stars as analysed in the sample of \cite{triana2017}. The same applies to the SOLA SAKs of \cite{dimauro2016}.
	
	\section{Conclusions}
	MOLA inversions have been shown to provide accurate estimates of the core rotation rates in red-giant stars. Gravity-dominated modes show a substantial amplitude in the core of red-giant stars that make the observed rotational splittings of these modes sensitive to its rotation. MOLA inversions efficiently suppress the sensitivity to the envelope when constructing core averaging kernels, which makes the kernels well-localised to the core region. However, surface rotation rates obtained with MOLA inversions are prone to substantial systematic errors \citep[e.g.,][]{ahlborn2020}. In contrast to the core averaging kernels, MOLA inversions are not able to suppress the sensitivity to the core region when constructing surface averaging kernels. As the cores of red giants generally rotate much faster than their envelopes \citep{beck2012,deheuvels2012,deheuvels2014,dimauro2016}, any sensitivity to the core region therefore substantially increases the estimate of the surface rotation rate.
	
	In this paper, we have shown that this deficiency of MOLA inversions can be attributed to the choice of the objective function. By integrating over the square of the averaging kernel, only its absolute value is taken into account, and a build-up of cumulative sensitivity is ignored. This works well for solar rotational inversions in which a large number of modes are available and oscillations within the individual rotational kernels do cancel each other properly. In red giants, the strongly-oscillating gravity-component in the core and the small number of observed modes render the complete cancellation of sensitivity in the core near impossible and prevent the surface averaging kernels from being more localised given the MOLA objective function. As long as the oscillations of the individual rotational kernels in the core have much smaller spatial scales than that on which the rotation profile changes, then they do not substantially alter the estimate of the rotation rate. The errors introduced by these small-scale oscillations are smaller than the systematic error introduced by the cumulative sensitivity to the core rotation. Hence, the surface averaging kernel should be constructed such that it more efficiently suppresses cumulative sensitivity in the core region.
	
	We have achieved this goal by introducing the extended MOLA objective function. In addition to the original term that integrates over the square of the averaging kernel, which suppresses large amplitudes away from the target radius, we extend the objective function with a second term taking the square of the integral over the averaging kernel, which suppresses the cumulative sensitivity away from the target radius. To maintain an optimal balance between both requirements, we have introduced a second trade-off parameter $\theta$ for the extension term. As with the trade-off parameter $\mu$, the new parameter also has to be calibrated. In Fig.~\ref{figcalib} we have shown how the cumulative sensitivity to the core region evolves for increasing $\theta$ for a core and a surface averaging kernel. The core sensitivity of the surface averaging kernel decreases monotonically with increasing $\theta$, crosses zero and reaches a negative asymptotic value for large $\theta$. At the zero-crossing, the trade-off parameter is optimal as the cumulative sensitivity is suppressed as intended. For the core averaging kernel, we observe a very similar behaviour. The sensitivity to the core region increases with increasing $\theta$ and also reaches an asymptotic value. In contrast to the surface averaging kernel, the core averaging kernel remains well-localised also for large values of the trade-off parameter. This means that a separate calibration of the trade-off parameter for the core averaging kernel is not required. Any positive value of $\theta$ will provide a better localised core averaging kernel than that obtained from MOLA inversions.
	
	The resulting cumulative core and surface averaging kernels are shown in Figs.~\ref{figcumuleMOLA} and \ref{figcumul}. As intended by construction, the surface averaging kernel does not show any cumulative sensitivity to the core region. Hence, the resulting surface rotation rate estimate will be an average value through the convective envelope independent of the core rotation. Similarly, the cumulative core averaging kernel rises very steeply in the core and therefore will provide an average rotation rate in the core region. As eMOLA gives an estimate of the surface rotation rate that is independent of the core, this will aid in drawing further conclusions about the rotation profile in the convective envelope when additional measures of the surface rotation are provided (e.g., from spectroscopy). When the rotation rate obtained from such additional measures are smaller than the estimate from asteroseismology, then this will be indicative that the rotation profile increases with depth in the convective envelope, for example like the convective power law profile. In the case where the rotation rate obtained from an additional measure agrees with the asteroseismic measure, the convective envelope most likely rotates as a solid body \citep{beck2018}. 
	
	To investigate this along a wider parameter space, we computed the inversion results along the lower part of the RGB. As for the model at the base of the RGB, we find that eMOLA inversions are capable of suppressing the sensitivity to the core rotation of the surface averaging kernel over the whole range of models investigated. For models both at the base and close to the bump of the RGB, estimates of the average envelope rotation rate can be obtained with the same level of accuracy. This offers the possibility to very accurately probe the average envelope rotation as a function of evolutionary state on the lower RGB. This will be a very important probe to further constrain the internal angular momentum transport in this evolutionary phase. For comparison, we also computed SOLA inversions along the lower part of the RGB. We find that while SOLA inversions work similarly well as eMOLA inversions for the less evolved stars, they face the same problems as MOLA inversions for the more evolved stars.
	
	We also compared our inversion results to those obtained with the approximation of linear rotational splittings proposed by \cite{goupil2013}. After obtaining the mode trappings from the displacement eigenfunctions of a suitable stellar model, core and surface rotation rates can be recovered with similar accuracy as eMOLA inversions. To interpret the results obtained from the linear fit, we computed the impact of each individual data point by means of linear regression. The linear regression coefficients allow for the construction of sensitivity kernels for the slope and the intercept in the same way as the inversion averaging kernels are constructed. Subsequently, we combined these sensitivity kernels linearly to obtain analogous averaging kernels for the estimators of core and envelope rotation rates. Both sensitivity kernels are very similar to the core and surface averaging kernels that we obtained with eMOLA inversions. This a posteriori comparison demonstrates why the estimated core and envelope rotation rates obtained from the linear rotational splittings are able to recover the input rotation very accurately and work similarly as eMOLA inversions. However, we would like to point out that in the eMOLA inversions the quality of the estimated rotation rates can be directly investigated and controlled by checking the localisation of the averaging kernels and tuning the inversion parameters. We therefore conclude that rotational inversions using eMOLA should be preferred, despite the simplicity of the LSA.
	\begin{acknowledgements}
		The research leading to the presented results has received funding from the European Research Council under the European Community’s Horizon 2020 Framework/ERC grant agreement no 101000296 (DipolarSounds). Funding for the Stellar Astrophysics Centre is provided by The Danish National Research Foundation (Grant agreement no.: DNRF106).
	\end{acknowledgements}
	
	%
	\bibliographystyle{bibtex/aa} 
	\bibliography{bibliography} 
	%
	\appendix
	\section{Derivation of eMOLA inversion coefficients}
	\label{secderiv}
	While the objective function of the extended MOLA inversions is different, the general principle of deriving the inversion coefficients is the same as for the original MOLA. The inversion coefficients are computed by minimising the objective function Eq.~(\ref{eqobj2}). The constraint of unimodular AKs is incorporated by including a Lagrange multiplier $\lambda$
	\begin{align*}
	Z^*= Z+2\cdot\lambda\left(\int_0^RK(r,r_0)\,\text{d} r-1\right).
	\end{align*}
	The factor of 2 which is introduced here makes the analytical derivation of the equations easier. To minimise the objective function $Z^*$ one evaluates:
	\begin{align*}
	\frac{\partial Z^*}{\partial c_i}=0\,\,\,\,\,\, \forall \,\,\,i,
	\end{align*}
	which provides an equation for each coefficient $c_i$. The equations for the coefficients are then written in matrix notation:
	\begin{align*}
	\mathbf{A}\cdot \vec{c}=\vec{v},
	\end{align*}
	where $\vec{c}$ contains all $c_i$ and $\lambda$ and $\vec{v}=(0,0,\dots,0,1)^\text{T}$. The matrix $\mathbf{A}$ is symmetric and is given as 
	\begin{align*}
	A_{ij}=\left\{
	\begin{array}{ll}
	A'_{ij}\,\,&i,j\leq M\\[5pt]
	\int_0^R\mathcal{K}_i(r)\,\text{d} r=\beta_i \,\,&i\leq M,j=M+1\\[5pt]
	\int_0^R\mathcal{K}_j(r)\,\text{d} r=\beta_j \,\,& j\leq M,i=M+1\\[5pt]
	0 \,\, &i,j=M+1
	\end{array}
	\right.
	\end{align*}
	with 
	\begin{align*}
	A'_{ij}=\;\theta&\int_0^R\mathcal{K}_i(r)J(r,r_0)\,\text{d} r\int_0^R\mathcal{K}_j(r)J(r,r_0)\,\text{d} r\\
	+&\int_0^R\mathcal{K}_i(r)\mathcal{K}_j(r)J(r,r_0)\,\text{d} r+\frac{\mu}{\mu_0} E_{ij}.
	\end{align*}
	The quadratic extension requires a modification of the inversion matrix $\mathbf{A}$ compared to MOLA inversions because all terms in the derivative of the modified objective function $Z$ (Eq.~(\ref{eqobj2})) still depend on the inversion coefficients. Irrespective of these modifications, the inversion coefficients are finally computed as:
	\begin{align}
	\vec{c}=\mathbf{A}^{-1}\cdot\vec{v}.
	\end{align}
	Due to the very similar structure of the inversion matrix, the implementation of this method into existing inversion codes can be achieved in a straightforward manner.
	
	\section{Choice of the trade-off parameter $\mu$}
\begin{figure}
		\centering
		\includegraphics[scale=1]{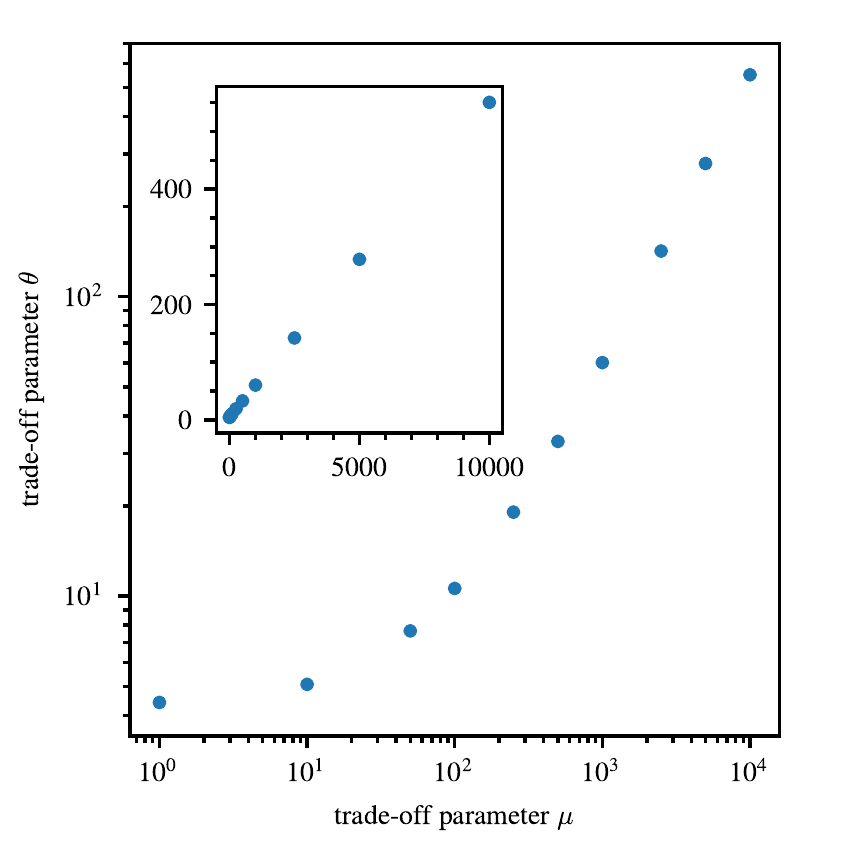}
		\caption{Calibrated trade-off parameter $\theta$ as a function of the trade-off parameter $\mu$. The inset shows the same quantities on a linear scale.}
		\label{figmu1}
	\end{figure}

\begin{figure}
		\centering
		\includegraphics[scale=1]{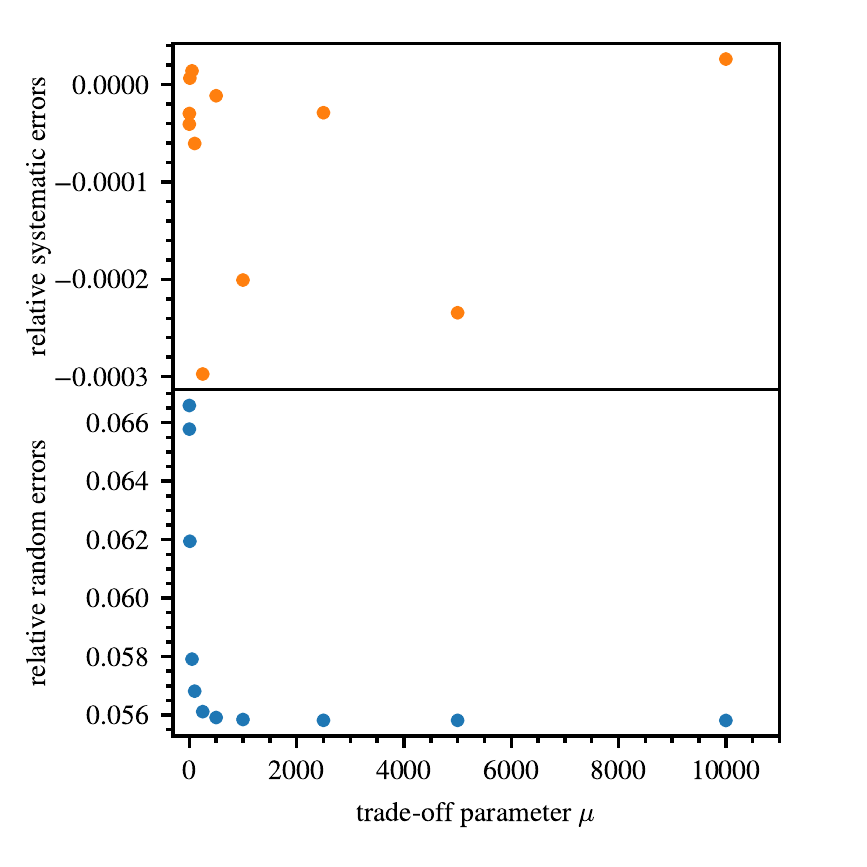}
		\caption{Relative random and systematic errors for a SAK as a function of the trade-off parameter $\mu$ for a calibrated value of $\theta$. Please note the different scales in the upper and lower panel.}
		\label{figmu2}
	\end{figure}

	In order to find a suitable value for the trade-off parameter $\mu$, we have calibrated the parameter $\theta$ for a range of different values $\mu$ as described in Sec.~\ref{sectheta}. The result is shown in Fig.~\ref{figmu1}. The optimal value of $\theta$ increases with increasing $\mu$. By increasing $\mu$ at fixed $\theta$, the localisation of the AK deteriorates, which can be counterbalanced by increasing $\theta$. The behaviour of the random and systematic errors for a SAK is shown in Fig.~\ref{figmu2}. As expected, the random errors go down with increasing $\mu$. They reach an asymptotic value around $\mu\approx1000$. At the same time, the systematic errors remain small. Please note that the scale of the systematic errors is lower than 0.1\%. Hence, the systematic errors are more than an order-of-magnitude smaller than the random errors. For the calibration of $\theta$ and the final rotational inversions, we have therefore chosen a value of ${\mu=1000}$. The same analysis can be done for the CAK. Here, a value of $\mu\approx1$ is sufficient to reduce the random errors while keeping the systematic errors small. The same value of $\theta$ as calibrated for the SAK has been used for the core.
	\label{secmu}
	\section{Systematic errors}
	\label{secsyserr}
\begin{figure}
		\centering
		\includegraphics{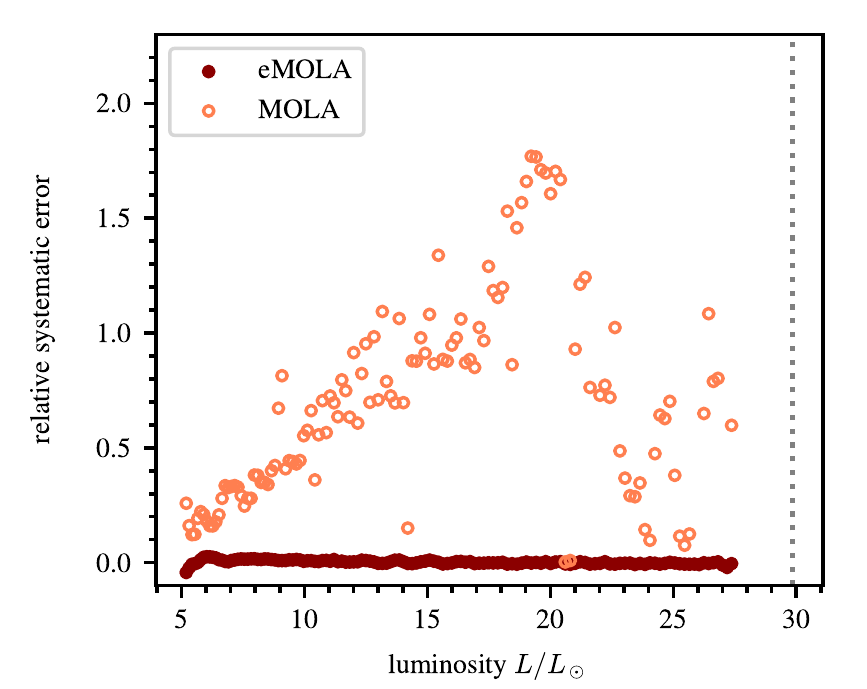}\\
		\includegraphics{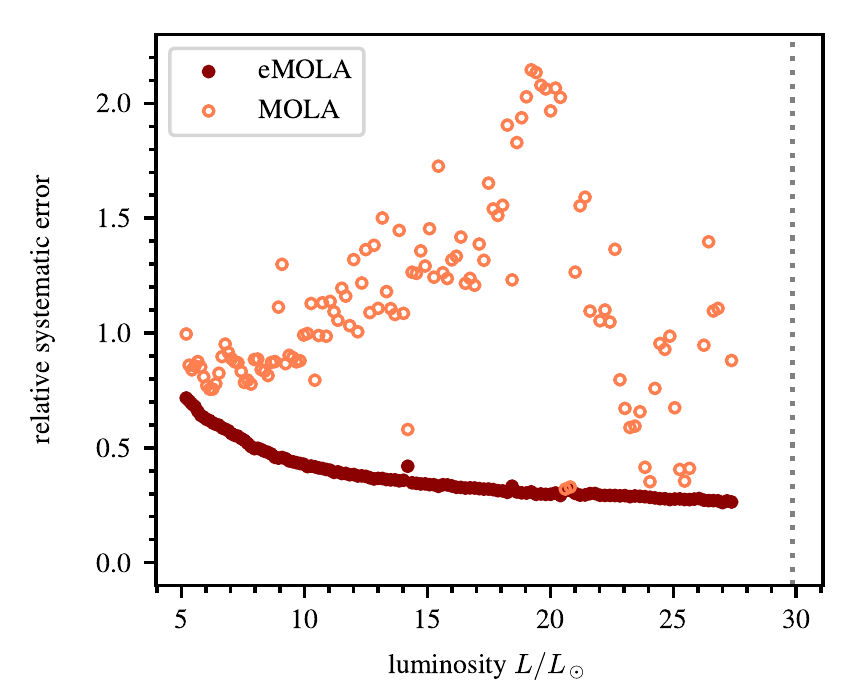}
		\caption{Relative systematic errors of the envelope rotation rates as a function of luminosity for MOLA and eMOLA inversions shown with light and dark red points, respectively. The bump luminosity is shown with vertical grey dotted line. 
			The upper panel shows the core step profile, in which eMOLA virtually eliminates the systematic errors that are present in the MOLA inversions.
			The lower panel shows the convective power law profile. The errors from eMOLA are generally much smaller than those from MOLA. 
		}
		\label{figsyserrcore}
	\end{figure}

	The trade-off parameter $\theta$ has been calibrated by minimising the sensitivity of the SAK below the base of the convection zone. In Fig.~\ref{figsyserr} we have demonstrated that this essentially eliminates all systematic errors when using the envelope step profile, as this profile has a step at the base of the convection zone. In this section, we show that the SAKs calibrated in this way work similarly well when assuming different stellar rotation profiles. In the upper panel of Fig.~\ref{figsyserrcore} we show the systematic errors of the estimated surface rotation rate when assuming the core step rotation profile as a function of the stellar luminosity. The systematic errors obtained from eMOLA inversions remain small along the whole evolutionary sequence and especially much smaller than the systematic errors obtained from MOLA inversions for the same underlying rotation profile. This shows that the SAKs computed with $\theta$ calibrated as described above are also able to recover the surface rotation when assuming a rotation profile with a step deeper inside the stellar core.
	
	In the lower panel of Fig.~\ref{figsyserrcore} we show the systematic errors of the estimated surface rotation rate when assuming the convective power law rotation profile. Consistent with the analysis of the envelope and core step rotation profiles, eMOLA provides an estimate of the average envelope rotation rate that is independent of the core rotation. Due to the rotation rate increasing with depth in the convective power law profile, the estimated average envelope rotation rate is larger than the value at the surface. Therefore, in contrast to the systematic errors obtained for the other profiles, the systematic errors of the estimated surface rotation rate for this profile amount to up to 70\% at the base of the RGB. However, as with the other two synthetic profiles, the systematic errors obtained from eMOLA inversions are smaller than those obtained from MOLA inversions. In addition, the eMOLA systematic errors decrease as the star evolves up the giant branch, allowing for more accurate estimates of the surface rotation rate in more evolved stars. We conclude that the parameter $\theta$ obtained with the adopted calibration procedure is also valid when assuming different stellar rotation profiles.
	\section{Comparison to SOLA inversions}
	\label{secSOLA}
	Another commonly used method for rotational inversions is the subtractive optimally localised averages (SOLA) inversion. The SOLA inversions have been first introduced by \cite{pijpers1992,pijpers1994}. In the past they have been used for solar rotational inversions \citep[e.g.][]{schou1998} and structural inversions \citep[e.g.][]{dziembowski1994,basu1996}, as well as for inversions in stars other than the Sun \citep[e.g.][]{reese2012,dimauro2016,bellinger2017,triana2017}. In contrast to MOLA inversions, the objective function is constructed by subtracting a target function from the AK instead of multiplying it. More specifically, the objective function of SOLA inversions is defined as:
	\begin{align}
	Z_{\rm SOLA}=\int_0^R\left[K(r,r_0)-T(r,r_0)\right]^2{\rm d}r+\frac{\mu}{\mu_0}\sum_{i,j\in\mathcal{M}}c_i(r_0)c_j(r_0)E_{ij}
	\end{align}
	where $K(r,r_0)$ is the AK and $T(r,r_0)$ is a target function. Minimising this objective function leads to a least-squares solution of the AK with respect to the target function. The target function is commonly defined as a Gaussian function centred at the target radius:
	\begin{align}
	T(r,r_0)=\frac{1}{f\Delta}\operatorname{exp}\left(-\left(\frac{r-r_0}{\Delta}\right)^2\right)
	\end{align}
	where $f$ is a normalisation constant to ensure a unit integral of $T(r,r_0)$. We would like to note that due to the limited amount of data and the mixed nature of all the oscillation modes observed in red-giant stars the resulting AKs will unlike in the solar case not resemble a Gaussian function at all. The inversion method will therefore most likely behave differently than in the solar case.
	
	By expanding the square parentheses the objective function can be rewritten to:
	\begin{align}
	Z_{\rm SOLA}=\int_0^R\left[K^2(r,r_0)-2K(r,r_0)\,T(r,r_0)+T^2(r,r_0)\right]{\rm d}r\nonumber\\
	+\frac{\mu}{\mu_0}\sum_{i,j\in\mathcal{M}}c_i(r_0)c_j(r_0)E_{ij}
	\label{eqobjSOLA}
	\end{align}
	The last term $T^2(r,r_0)$ of the integral is independent of the inversion coefficients $c_i$. For a given $\Delta$, it will just shift the objective function up and down without changing the location of the minimum. It will hence drop out when deriving with respect to the coefficients. We therefore neglect this term and limit our further discussion to the first two terms. For the sake of simplicity, we set $\mu=0$ as well. Neglecting the last term, the objective function reads:
	\begin{align}
	Z^*_{\rm SOLA}=\int_0^RK^2(r,r_0)\,{\rm d}r-\int_0^R2K(r,r_0)\,T(r,r_0)\,{\rm d}r
	\label{eqobjSOLAred}
	\end{align}
	Hence, minimising this objective function $Z^*_{\rm SOLA}$, which is equivalent to minimising $Z_{\rm SOLA}$ Eq.~(\ref{eqobjSOLA}), means to minimise two competing terms. The first term looks like it is minimising the overall amplitude of the kernel, e.g. suppressing oscillatory solutions. This term is similar to the first term of the MOLA objective function, Eq.~(\ref{eqobjMOLA}), without the function $J(r,r_0)$. Due to the minus sign, the second term is maximising the sensitivity of the AK within the target function $T$. Due to the normalisation of the AKs maximising the sensitivity within a region is equivalent to minimising the sensitivity outside a certain region. This is exactly the idea of the second term we have introduced in the eMOLA objective function Eq.~(\ref{eqobj2}). 
	
	Changing the inversion parameter $\Delta$ changes the behaviour of the second term by changing the target function. It especially changes the ratio of the first to the second term of the SOLA objective function, Eq.~(\ref{eqobjSOLAred}). However, it has a rather indirect impact on the ratio, as it only acts through a change of the objective function and does not directly scale the second term. As discussed above, scaling the second term is the role taken by the eMOLA trade-off parameter $\theta$ in Eq.~(\ref{eqobj2}). This property is important to understand the behaviour of the SOLA inversions as a function of evolution.
	\begin{figure}
		\centering
		\includegraphics[scale=1]{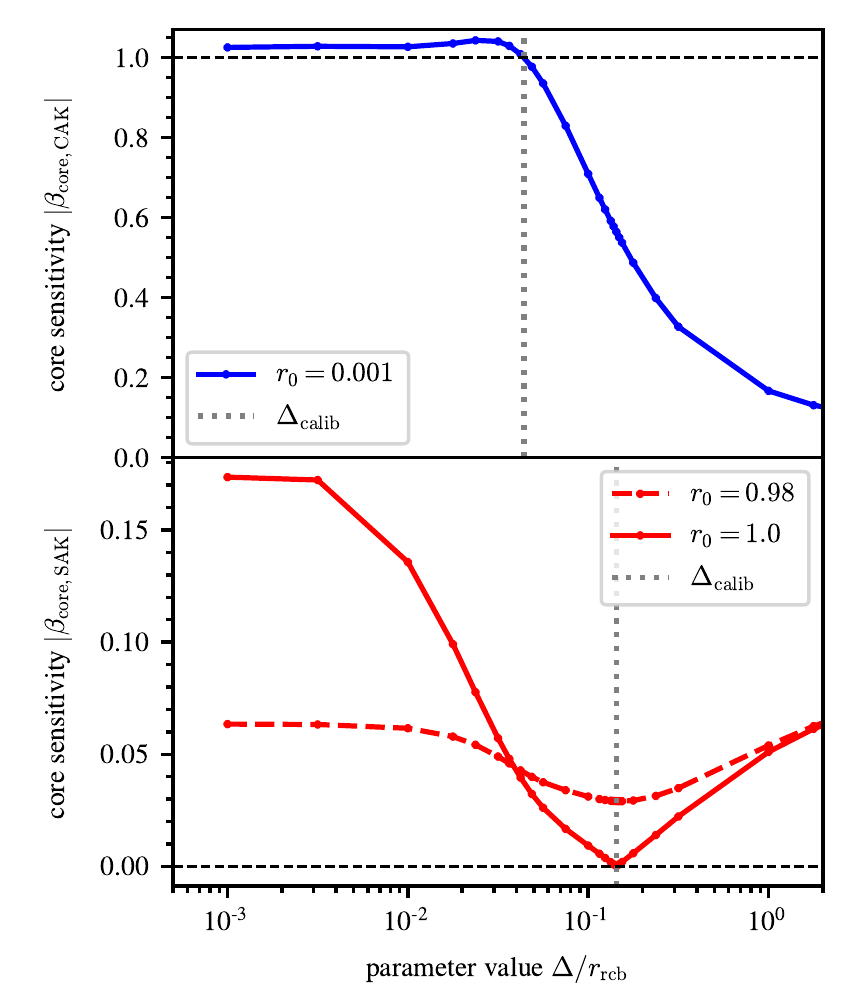}
		\caption{Absolute value of the core sensitivity $|\beta_{\rm core}|$ as a function of the target function width $\Delta$. The results for core and surface averaging kernels are shown in the upper and lower panel, respectively. The core averaging kernel has a target radius of $r_0/R=0.001$ while the surface averaging kernels have target radii of $r_0/R=0.98$ and 1.0. The optimal value of $\Delta$ for $r_0/R=0.001$ and 1.0 are shown with the vertical dotted line in each panel.}
		\label{figdelta1}
	\end{figure}
	The parameter $\Delta$, which defines the width of the target function, needs to be calibrated, as the parameters of the other inversion methods. Here, we adopted a similar approach as for the trade-off parameter $\theta$ and as was outlined in \cite{triana2017}. This is illustrated in Fig.~\ref{figdelta1} in which we show the absolute value of the core sensitivity $|\beta_{\rm core}(r_0)|$ as a function of $\Delta$. The upper panel shows the calibration procedure for the CAK. The width of the target function is considered optimal when the core sensitivity of the CAK is as close to one as possible. The result of the calibration is indicated with a vertical grey dotted line. The calibration for a SAK is shown in the lower panel. For this case, the width of the target function is optimised such that the core sensitivity is minimised. The kink in the results for $r_0/R=1.0$ occurs as the core sensitivity changes sign. As for the eMOLA inversions, this can be attributed to the occurrence of negative side lobes. For $r_0/R=0.98$, it is not possible to find a parameter $\Delta$ that completely removes the sensitivity to the core from the SAK.
	
	\begin{figure}
		\centering
		\includegraphics[scale=1]{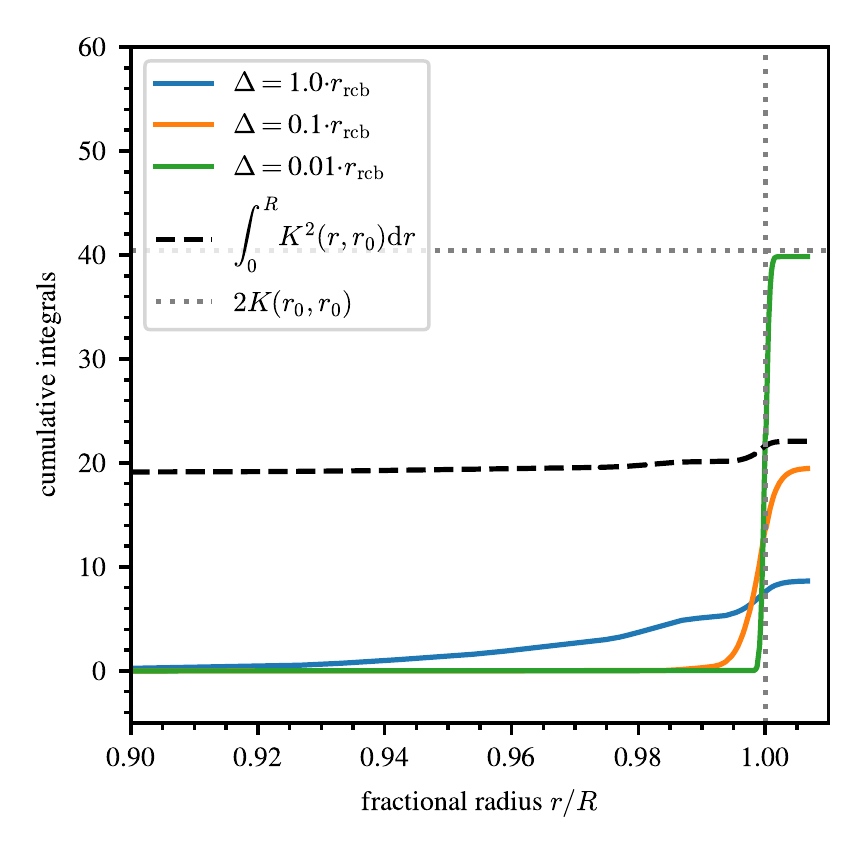}
		\caption{Cumulative integral of the first and second term of the SOLA objective function Eq.~(\ref{eqobjSOLA}) as a function of fractional radius for different values of $\Delta$. The first term is indicated with the black dashed line, while the second term is indicated with the coloured lines. The horizontal grey dotted line indicates the asymptotic value of the second term for $\Delta\to 0$.}
		\label{figdelta2}
	\end{figure}
	
	To understand how varying the width of the target function allows us to optimise the sensitivity of the AKs in certain regions, we will further discuss the two terms of the objective function Eq.~(\ref{eqobjSOLAred}) for different values of $\Delta$. The behaviour of the objective function is illustrated in Fig.~\ref{figdelta2}. For the sake of simplicity, we assume a fixed AK for our discussion of varying $\Delta$. For real inversions, the AKs would of course change when varying $\Delta$. The coloured lines show the cumulative integral of the second term of $Z^*_{\rm SOLA}$ as a function of radius for a fixed AK with target-radius $r_0/R=1.0$ and target functions with varying width $\Delta$. The black dashed line shows the cumulative integral of the first term as a function of radius that is independent of $\Delta$ for a given AK. For large values of $\Delta$, the second term has a smaller magnitude than the first term. For decreasing the width of the target function, the magnitude of the second term increases further and further and eventually surpasses the magnitude of the first term. Asymptotically, for $\Delta\to 0$, the target function transitions into a Dirac-delta distribution $\delta(r-r_0)$. For the second term, one can hence see that
	\begin{align}
	\lim_{\Delta\to0}\int_0^R2K(r,r_0)T(r,r_0){\rm d}r=2K(r_0,r_0)
	\end{align}
	In Fig.~\ref{figdelta2} we show this limit value with a horizontal grey dotted line. For finite values of $\Delta$, the integral remains finite. This shows immediately that the second term cannot grow arbitrarily. Depending on the data given and the morphology of the AK close to the surface, one might encounter situations in which the first term dominates the objective function Eq.~(\ref{eqobjSOLAred}). A dominating first term leads however to badly localised AKs, as discussed for the MOLA objective function. In contrast to the target function width $\Delta$, the trade-off parameter $\theta$ of the eMOLA objective function which we introduced in front of the second term in Eq.~(\ref{eqobj2}) is able to scale the second term to arbitrarily large values and ensures the optimal balance of the first to the second term. As described in \cite{ahlborn2020} the sensitivity of the rotational kernels to the core increases as the stars evolve up the RGB. It is hence expected that for more evolved stars, the first term of the SOLA objective function Eq.~(\ref{eqobjSOLA}) becomes more important. If it increases enough to dominate the behaviour of the objective function, the SOLA inversions behave similar to the MOLA inversions and also SOLA SAKs show cumulative sensitivity to the core region. Further, the target radius needs to be chosen with care. If the target radius coincides with a local minimum in the SAK the magnitude of the second term is reduced and it might happen so that even for less evolved stars the first term of the objective function dominates such that the cumulative sensitivity to the core region cannot be removed. This behaviour is illustrated in the lower panel of Fig.~\ref{figdelta1}. While it is possible to remove the cumulative core sensitivity when choosing a target radius of $1.0$, some cumulative core sensitivity remains when choosing $r_0/R=0.98$, which has been used as the default target radius for the eMOLA inversions.
	
\begin{figure}
		\centering
		\includegraphics[scale=1]{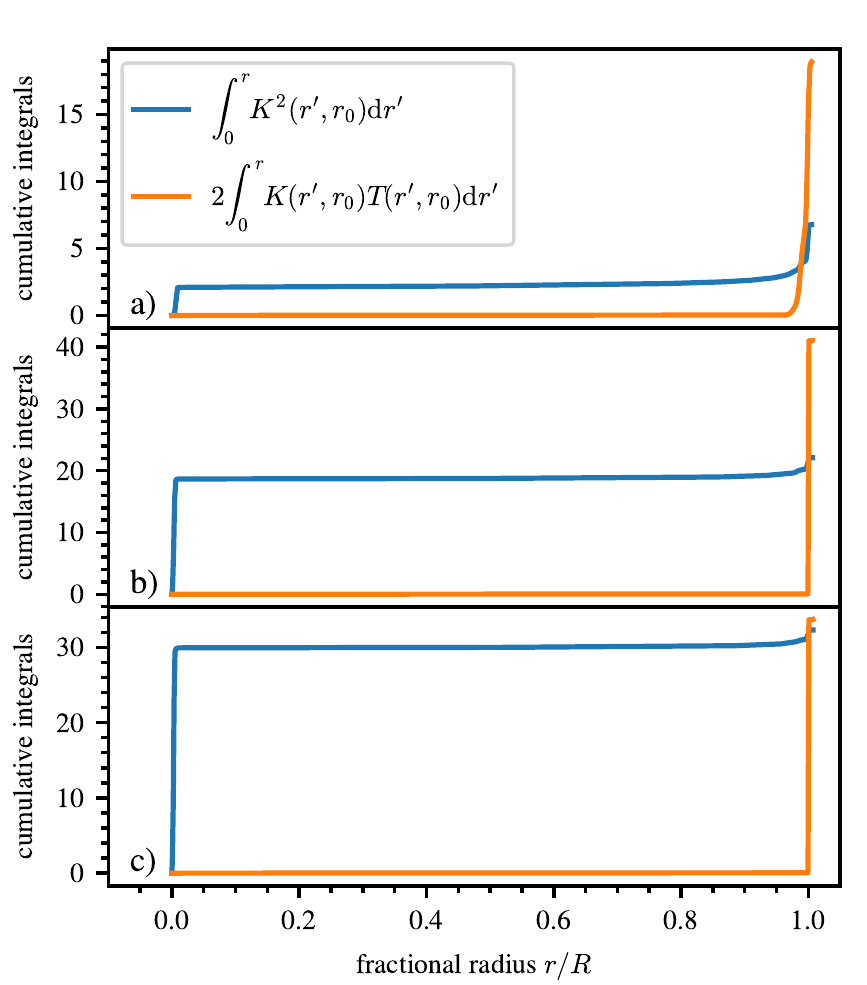}
		\caption{Cumulative integral of the first and second term of the SOLA objective function Eq.~(\ref{eqobjSOLA}) for optimal values of $\Delta$ as a function of fractional radius shown with a blue and orange line, respectively. Panel a) corresponds to the model discussed in Sec.~\ref{seceMOLAsingle}, panel b) corresponds to a model around 15~$L_\odot$ and panel c) corresponds to a model at about 20~$L_\odot$, i.e. is right at the maximum of systematic errors in Fig.~\ref{figsyserr}.}
		\label{figcomponents}
	\end{figure}

	After discussing the behaviour of the SOLA target function for a single stellar model, we will now discuss how the objective function Eq.~(\ref{eqobjSOLA}) changes as the stars evolve up the RGB. In Fig.~\ref{figcomponents} we show the cumulative integrals of the two components of the objective function $Z^*_{\rm SOLA}$ Eq.~(\ref{eqobjSOLAred}) for optimal values of $\Delta$ as a function of fractional radius. The three panels refer to stellar models in different evolutionary stages, where the first panel is the least evolved model and the last panel is the most evolved model. In the first panel, the first term gains about half of its value in the deep interior and the other half close to the surface. Further, its final value is less than half the value of the second term. As the star evolves, two things happen: (i) the first term gains more and more of its value in the deep interior and (ii) the ratio of the first to the second term increases. This can be explained by the increase in  sensitivity to the core region in the individual rotational kernels. In panel c) it is apparent that the first term gains nearly all of its value in the core and reaches about the same value as the second term close to the surface. As discussed above, the magnitude of the second term cannot be increased arbitrarily. This is the reason why the SOLA inversion is not able to suppress cumulative sensitivity to the core in the SAKs for the more evolved stars.
	
	This discussion shows that applying SOLA inversions to red-giant stars is facing some challenges, that mostly occur due to the mixed nature of the oscillation modes and the limited amount of data available. Even though SOLA inversions work for less evolved red giants, this result is numerically not robust, as already a small change of the target-radius changes the properties of the resulting AK substantially. Further, SOLA inversions are not able to recover accurate surface rotation rates for more evolved stars. The inversion parameter $\Delta$ is changing the objective function only in an indirect way, which limits the tuning potential of this parameter. As discussed above, the ability to tune the parameter to obtain accurate results vanishes for an increasing core sensitivity of the oscillation modes. In contrast to that, eMOLA inversions are robust to small changes of the target radius and the inversion parameter $\theta$ can be tuned also for the models which are most sensitive to the core region.
\end{document}